\documentclass[aip,jmp,amsmath,amssymb,reprint]{revtex4-1}

\usepackage{graphicx}
\usepackage{dcolumn}
\usepackage{bm}
\usepackage[centering,hmargin=2cm,vmargin=2cm]{geometry}

\begin{document}

\title{Proton network flexibility enables robustness and large electric fields in the ketosteroid isomerase active site}

\author{Lu Wang}
\affiliation{Department of Chemistry and Chemical Biology, Rutgers University, Piscataway, New Jersey 08854, USA}

\author{Stephen D. Fried}
\affiliation{Medical Research Council Laboratory of Molecular Biology, Francis Crick Avenue, Cambridge CB2 0QH, UK}

\author{Thomas E. Markland}
\email{tmarkland@stanford.edu}
\affiliation{Department of Chemistry, Stanford University, Stanford, California 94305, USA}

\date{\today}

\begin{abstract}
Hydrogen bond networks play vital roles in biological functions ranging from protein folding to enzyme catalysis. Here we combine electronic structure calculations and {\it ab initio} path integral molecular dynamics simulations, which incorporate both nuclear and electronic quantum effects, to show why the network of short hydrogen bonds in the active site of ketosteroid isomerase is remarkably robust to mutations along the network and how this gives rise to large local electric fields. We demonstrate that these properties arise from the network's ability to respond to a perturbation by shifting proton positions and redistributing electronic charge density. This flexibility leads to small changes in properties such as the partial ionization of residues and $pK_a$ isotope effects upon mutation of the residues, consistent with recent experiments. This proton flexibility is further enhanced when an extended hydrogen bond network forms in the presence of an intermediate analog, which allows us to explain the chemical origins of the large electric fields in the enzyme's active site observed in recent experiments.
\end{abstract}

\maketitle

\twocolumngrid

\section{Introduction}
Hydrogen bonds play essential roles in biological systems, carrying out diverse functions ranging from stabilizing protein structures, to modulating molecular recognition and catalyzing enzymatic reactions.\cite{Pauling1951,Dill1990,Nelson2008,Zhang2016,Cui2016} An important class of these are low-barrier hydrogen bonds (LBHBs), which can occur when the donor-acceptor distance between heavy atoms, $R$, is below 2.6 \AA~and the proton affinities of the hydrogen bonding partners are closely matched.\cite{Hibbert1990,Cleland1994,Tuckerman1997,Perrin1997,Cleland1998,marx06cpc,mckenzie2012,McKenzie2014} LBHBs have been implicated in a wide variety of biological processes such as catalyzing chemical reactions\cite{Gerlt1993,Cleland1998}, promoting structural stability\cite{Yamaguchi2009} and facilitating proton transfer.\cite{Pinotsi2016} LBHBs are more commonly observed in proteins than in liquids as the protein's three-dimensional fold can help position the residues closer together. While the values for $R$ in LBHBs are only slightly shorter than typical hydrogen bonds in liquids, such as water where $R \sim$ 2.8 \AA, this shortening significantly alters the potential energy surface for transferring the proton between the donor and acceptor. As an illustration, consider an O--H bond with a typical length of 1 \AA. When $R$ = 2.8 \AA, the distance between the two minima, arising from the H being bound to the donor or to the acceptor, in the potential curve is $\sim$0.8 \AA. Decreasing $R$ to 2.6 \AA~shortens this distance to $\sim$0.6 \AA. This $\sim$25\% decrease brings the distance between the potential energy minima close to the proton's thermal wavelength. In these cases the proton transfer barrier can become comparable to the zero point energy (ZPE) for the Donor--H bond (typically $\sim$5 kcal/mol), allowing the proton to delocalize between the donor and acceptor.

\begin{figure}[ht]
\centering
\includegraphics[width=\columnwidth]{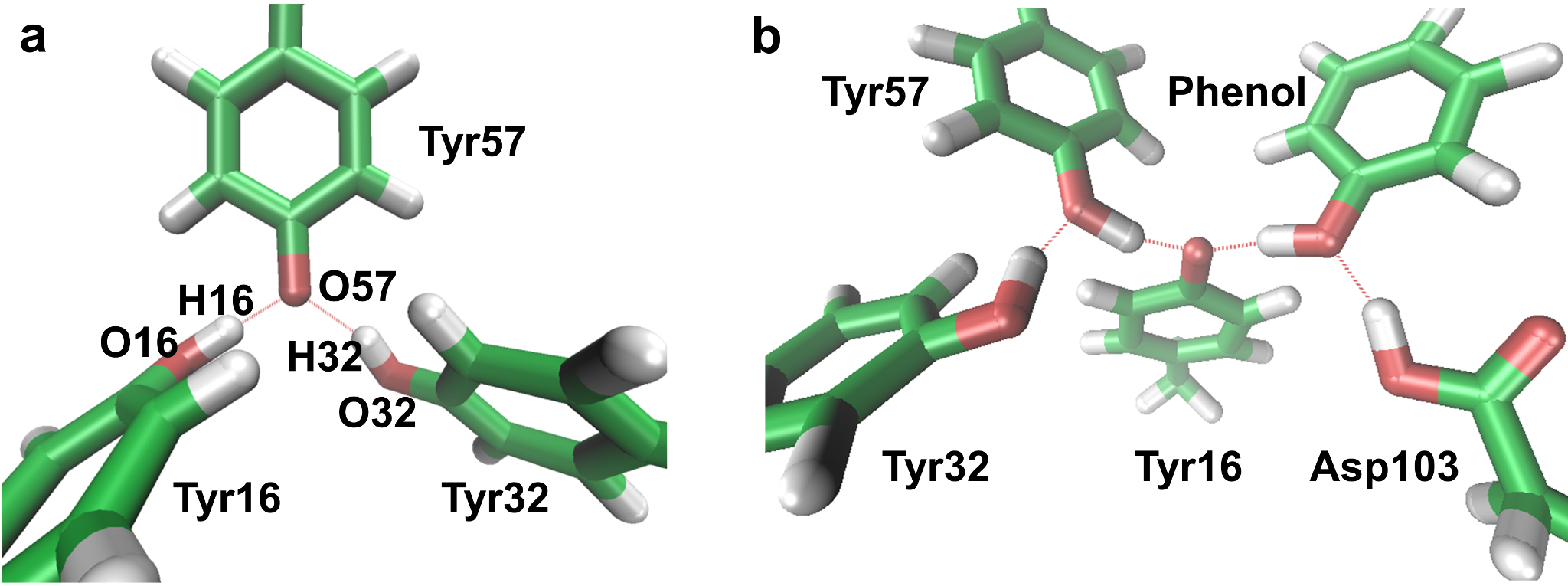}
\caption{Hydrogen bond network in the active site of KSI$^{D40N}$. (a) Local network formed from residues Tyr16, Tyr32 and Tyr57 with a triad structural motif, with the hydrogen bonded oxygen and hydrogen atoms labeled. (b) Extended network formed in the presence of an intermediate analog phenol. Red, green and white represent oxygen, carbon and hydrogen atoms, respectively.}
\label{fig:triad}
\end{figure}

While many biological systems exploit single hydrogen bonds in their functions, others invoke networks of hydrogen bonds.\cite{Murgida2001,marx06cpc,Ball2008,Reece2009,Pinotsi2016} The enzyme ketosteroid isomerase (KSI) provides one such intriguing system, which contains a network of short hydrogen bonds in its active site.\cite{Ha2000,Kraut2006,Sigala2013} KSI catalyzes the isomerization of steroid substrates with a 10$^{11}$-fold rate enhancement compared to the analogous reaction in solution.\cite{Pollackl1989,Hawkinson1991,Radzicka1995,Pollack2004,Warshel2007,Kamerlin2010,Chakravorty2010} Its catalytic mechanism involves proton transfer from the substrate to the enzyme to form a negatively charged intermediate, which is stabilized by the enzyme's hydrogen bond network. This is followed by proton donation from the enzyme back to the substrate at a different position, leading to a net isomerization of the substrate (Fig. S1).\cite{Pollack2004,Herschlag2013}  
KSI's ability to stabilize its reaction's intermediate through the active-site hydrogen bond network can be investigated using its mutant, KSI$^{D40N}$, which preserves the structure of the wild-type enzyme and mimics its ionization state in the enzyme-intermediate complex.\cite{Zhao1996,Ha2000,Kraut2006,Childs2010,Sigala2013} In the absence of an intermediate analog, the side chain of residues Tyr16, Tyr32 and Tyr57 in KSI$^{D40N}$ form a hydrogen bonded triad with $R \sim$ 2.6 \AA, as shown in Fig. \ref{fig:triad}a.\cite{Ha2000,Sigala2013} This network of LBHBs promotes quantum proton delocalization and facilitates the deprotonation of Tyr57, leading to a very high acidity of this residue ($pK_a$=5.8;\cite{Wu2015} tyrosine in aqueous solution has a $pK_a$ of 10.1) and a prominent H/D isotope effect  ($\Delta pK_a$=0.9;\cite{Wu2015} typical organic acids have a $\Delta pK_a$ of 0.4-0.6\cite{Wolfsberg2009}). Upon binding of an inhibitor, an extended network of short hydrogen bonds is formed to incorporate the inhibitor and residue Asp103 (Fig. \ref{fig:triad}b).\cite{Kraut2006,Hanoian2010,Chakravorty2010,Sigala2013,Wang2014a} KSI$^{D40N}$ thus provides an ideal system to assess the behavior of networks of LBHBs in a biological system. The donor-acceptor distances, $R$, are short and the $pK_a$'s of the hydrogen bond participants in solution are closely matched, suggesting that they share similar proton affinities in the network. These are generally considered as two key prerequisites for forming LBHBs.\cite{Hibbert1990,Perrin1997,Cleland1998,mckenzie2012}

\begin{figure}[ht]
\centering
\includegraphics[width=\columnwidth]{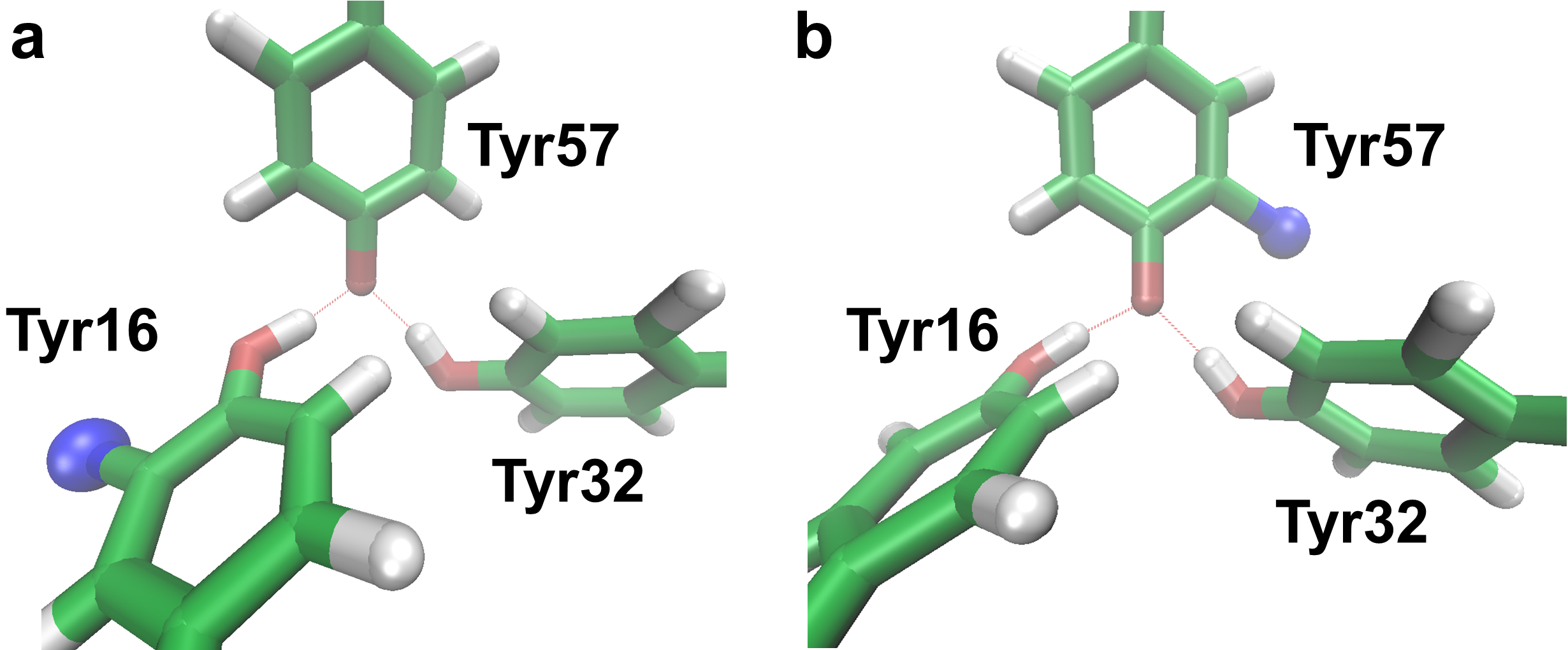}
\caption{Perturbations to the active-site hydrogen bond network of KSI$^{D40N}$: (a) Tyr16 replaced by 3-chlorotyrosine and (b) Tyr57 replaced by 3-chlorotyrosine. Red, green, blue and white represent oxygen, carbon, chlorine and hydrogen atoms, respectively.}
\label{fig:label}
\end{figure}

Recent experiments have revealed a number of intriguing observations about how KSI responds to chemical perturbations in its active site.\cite{Schwans2013,Natarajan2014,Fried2014,Wu2015,Wu2016,Lamba2016,Fried2017} In particular, mutagenesis experiments have been conducted that employ unnatural amino acids to perturb $R$ and the proton affinities of the active-site residues while keeping the overall hydrogen bond network intact. These studies have shown that KSI's catalytic rate and the properties of the network are remarkably robust with respect to such perturbations. For example, a recent experiment substituted Tyr16 with a series of fluorotyrosines and found that, despite the fact that fluorination results in a 40-fold increase in tyrosine's acidity in aqueous solution, only a 1.2-fold decrease in KSI's catalytic rate constant was observed.\cite{Natarajan2014} In another study, the tyrosine residues in the triad were substituted with their chlorinated counterparts (Fig. \ref{fig:label}a and b), which led to only minor changes in properties such as the partial ionization and H/D isotope effects.\cite{Wu2015} In addition, recent vibrational Stark effect spectroscopy experiments used a product-like inhibitor that binds to wild-type KSI and identified a very large electric field of -144$\pm$6 MV/cm in the enzyme's active site (compared to -80$\pm$2 MV/cm in water).\cite{Fried2014} The strength of the fields was demonstrated to strongly correlate with the enzyme's catalytic rate enhancement and hence it was suggested that they may play a role in the enzyme's function.\cite{Fried2014,Fried2017} However, despite the apparent importance of these fields, previous studies invoking both point charge and polarizable empirical potentials have observed average electric fields of less than half the experimental value.\cite{Fried2014,Bhowmick2017}

Here, we use {\it ab initio} simulations to analyze the quantum fluctuations of the electrons and nuclei in the active-site hydrogen bond network of KSI$^{D40N}$ and uncover the origins of the robustness of the network in response to chemical perturbations. We then show that extending the size of the hydrogen bond network by binding an intermediate analog (Fig. \ref{fig:triad}b)\cite{Sigala2013} further enhances the movement of protons and electrons in the active site, resulting in the large electric fields acting on the intermediate analog. Finally, we discuss the possible biological implications of the ability of the extended hydrogen bond to respond to perturbations arising from mutations in the network. 

\section{Simulation methods}
To examine the role of structural fluctuations, we carried out {\textit{ab initio}} molecular dynamics (AIMD) simulations via an MPI interface\cite{Isborn2012} to the TeraChem program\cite{Ufimtsev2009}. Since nuclear quantum effects (NQEs) are important in hydrogen bonded systems,\cite{chen+03prl,morr-car08prl,Perez2010,li+11pnas,mark-bern12pnas,ceri+13pnas,Wang2014a,Wang2014,McKenzie2015,Rossi2015,Pinotsi2016,Ceriotti2016} we also examined the effect of NQEs using \textit{ab initio} path integral molecular dynamics (AI-PIMD) simulations, which treat both the nuclei and electrons quantum mechanically.\cite{feyn-hibb65book,chan-woly81jcp,bern-thir86arpc} 

AIMD and AI-PIMD simulations were performed using a QM/MM setup for the following systems: KSI$^{D40N}$ (initial structure from PDB 1OGX),\cite{Ha2000}
KSI$^{D40N}$ with Tyr16 or Tyr57 replaced by 3-chlorotyrosine (initial structures from PDB 5D82 and 5D81,\cite{Wu2015} respectively), KSI$^{D40N}$ with a bound phenol (initial structure modified from PDB 3VGN\cite{Sigala2013}) and tyrosine in aqueous solution. Additional simulation details are given in the Supporting Information. We have also summarized these simulations methods in a recent review.\cite{Wang2016}

Partial ionizations of residues were calculated using a linear interpolation scheme, which is consistent with the method to extract ionization values from $^{13}$C NMR experiments.\cite{Wang2014,Wu2015} To justify the scheme, we calculated $^{13}$C NMR chemical shifts of Tyr16's C$_\zeta$ with the proton H16 moving along the O16--O57 vector. As shown in Fig. S2, the calculated $^{13}$C chemical shifts have a strong linear dependence on the O--H bond length. For each hydrogen bonded pair O--H--O$^\prime$, we define the atom O to be completely neutral if the distance between O and H, $d_{OH} \leq 0.96$ \AA~and it to be completely ionized if $d_{O^\prime H} \leq 0.96$ \AA. The amount of partial ionization is obtained by linearly interpolating the two limits. We note that this definition is different from the one we exploited previously, where we used the proton transfer coordinate $\nu=d_{OX,HX}-d_{O57,HX}$ (X=16 or 32) and defined a tyrosine to be ionized when $\nu\geq 0$ for H16 or H32.\cite{Wang2014a} 

The excess isotope effect, $\Delta\Delta pK_a$ (the difference between the active site tyrosine’s H/D pKa isotope effect and that of tyrosine in water), was calculated from the free energy differences using the thermodynamic cycles shown in Fig. S3. $\Delta\Delta pK_a$ was further decomposed into three orthogonal coordinates, one along the O--H bond direction, one in the plane of the O--H--O$^\prime$ hydrogen bond but perpendicular to the O--H bond vector, and one perpendicular to the O--H--O$^\prime$ hydrogen bond plane.
The decomposition was calculated by projecting the free energies along the three directions.  

The active-site electric fields in KSI$^{D40N}$, $E_{CO}$, were calculated as the electric fields exerted by the active-site atoms on the C--O chemical bond of the phenol molecule, projected along the C--O bond unit vector. We extracted 6142 complex configurations from the AI-PIMD simulations of KSI$^{D40N}$ with phenol bound, with a time spacing between configurations of 5 fs. Each complex configuration contains Tyr16, Tyr32, Tyr57, Asp103 and phenol, and we evaluated the total electric fields on the C and O atoms in the phenol molecule, which were then projected along the C--O bond direction. The total electric field on the C--O bond in the enzyme complex, $E_{complex}$, was calculated by averaging over the fields on the C and O atoms. To remove the contribution from the phenol molecule itself, $E_{phenol}$ (i.e., the self-field), we repeated the calculation on the deprotonated phenol molecule as isolated from each complex configuration. $E_{CO}$ was obtained by $E_{complex}-E_{phenol}$. The calculations were performed using the Gaussian 09 program\cite{g09} using the B3LYP functional\cite{beck93jcp} and 6-31G* basis set.   We did not include the protein environment in these calculations. To justify this, we extracted 422 configurations from AI-PIMD simulations, and calculated $E_{CO}$ with and without a point charge environment from the rest of the protein, water and counter-ions (partial charges according to the AMBER03 force field\cite{Duan2003} and TIP3P water model\cite{jorg+83jcp}). The resulting difference in electric fields was less than 2\%.

\section{Results and discussion}

By incorporating unnatural amino acids to perturb the active-site hydrogen bond network in KSI$^{D40N}$, recent experiments have shown that this network is surprisingly robust in that the ionization states of the residues and the $pK_a$ isotope effects are almost unchanged upon making these perturbations.\cite{Wu2015} After binding an inhibitor, Stark effect spectroscopy measurement were used to show that the active site of the enzyme exerts a large electric field on a bond of the inhibitor, which undergoes a charge redistribution during the catalytic cycle.\cite{Fried2014,Fried2017} Here, we discuss these properties in turn and elucidate how the delocalization of the protons in the hydrogen bond network gives rise to these observations.

\subsection{Change in partial ionizations in response to mutations}

We begin by considering the effect of chlorination on the hydrogen-bonded triad of tyrosine residues in the active site of KSI$^{D40N}$. In the absence of mutations (Fig.~\ref{fig:triad}a), the triad motif formed in KSI$^{D40N}$ allows the protons H16 and H32 to be quantum mechanically delocalized in the network.\cite{Wang2014a} This delocalization causes the side-chain phenol groups of the three tyrosine residues to be partially ionized. Equilibrium partial ionizations of residues Tyr16, Tyr32 and Tyr57 were obtained from $^{13}$C NMR experiments assuming a linear change in the NMR chemical shift of the carbon adjacent to the ionizable oxygen ($^{13}$C$_\zeta$) between the ionized and unionized state, and yielded values of 20\%, 13\%, and 67\%, respectively.\cite{Wang2014a,Wu2015} Our electronic structure calculations exhibit a close-to-linear relationship between the calculated $^{13}$C NMR chemical shift and the position of the proton when it is moved between the hydrogen bond donor and acceptor (Fig. S2) giving \emph{post hoc} justification to this experimental assumption. Using this relation, our AI-PIMD simulations predict that residues Tyr16, Tyr32 and Tyr57 are 14\%, 13\% and 73\% ionized, respectively, in very good agreement with experiment. 

We now consider two mutants, which have been investigated in recent experiments\cite{Wu2015}, where either Tyr16 or Tyr57 is replaced by 3-chlorotyrosine. The resulting proteins are referred to as Cl-Tyr16 and Cl-Tyr57 KSI$^{D40N}$, which are depicted in Fig. \ref{fig:label}a and 2b, respectively. In aqueous solution, chlorination lowers the side-chain $pK_a$ of tyrosine by 1.8 units,\cite{Wu2015} which is equivalent to stabilizing the deprotonated state by 2.5 kcal/mol. If one were to assume a simple two-state model to describe the proton transfer between Tyr16 and Tyr57, i.e., H16 is attached to either one tyrosine or the other (Fig. \ref{fig:triad}a), then one would expect chlorination of Tyr16 ($\delta_{pK_a}=1.8$) to increase the relative ionization of Tyr16 to Tyr57, denoted as $f_{16}/f_{57}$, by $e^{2.303\delta_{pK_a}}=63$ fold compared to the unlabeled protein. In addition, a key assumption of LBHBs is that the proton affinities of the hydrogen bonding partners are closely matched. As chlorination breaks the matching of the solution $pK_a$'s of the residues in the triad, one might expect a large change in the properties of the triad \cite{Hibbert1990,Cleland1994,Perrin1997,Cleland1998}.

In contrast to this prediction, our AI-PIMD simulations predict that chlorinating the tyrosine residues only slightly affects the ionization states in the triad. As shown in Table \ref{tab:partial}, in Cl-Tyr16 KSI$^{D40N}$ $f_{16}/f_{57}$ is increased by 1.34-fold compared to the unlabeled protein, which is in excellent agreement with the experimentally measured value of 1.45-fold.\cite{Wu2015} In contrast, chlorinating Tyr57 reduces proton sharing in the triad, which decreases $f_{16}/f_{57}$ by 1.2 fold compared to the unlabeled case. 
\begin{table}
\resizebox{0.9\columnwidth}{!}{
\begin{tabular}{cc|ccc}
\hline\hline
& Unlabeled & $\Delta$(Cl-Tyr16) & $\Delta$(Cl-Tyr57) \\
\hline
Tyr16   &  14\% (20\%) & +4\% (+7\%)  & -1\%  \\
Tyr32   &  13\% (13\%) & -1\% (-2\%)  & -5\%   \\
Tyr57   &  73\% (67\%) & -3\% (-5\%)  & +6\%  \\
\hline\hline
\end{tabular}
}
\caption{Partial ionizations of residues Tyr16, Tyr32 and Tyr57 in the unlabeled KSI$^{D40N}$ and their changes in Cl-Tyr16 and Cl-Tyr57 KSI$^{D40N}$ compared to the unlabeled protein ($\Delta$(Cl-Tyr16) and $\Delta$(Cl-Tyr57) respectively). Also included in brackets are the values obtained from experiments.\cite{Wu2015}}
\label{tab:partial}
\end{table}

\begin{figure*}[ht]
\centering
\includegraphics[width=2\columnwidth]{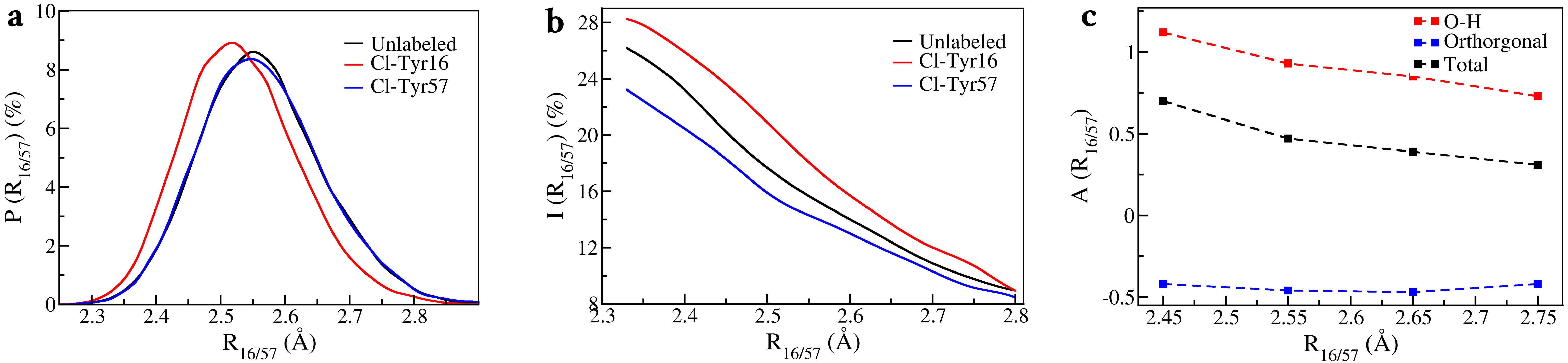}
\caption{The distribution of $R_{16/57}$ (a), the ionization propensity $I (R_{16/57})$ (b), and the $pK_a$ isotope effect $A (R_{16/57})$ (c) for unlabeled, Cl-Tyr16 and Cl-Tyr57 KSI$^{D40N}$.}
\label{fig:triad_decomp}
\end{figure*}

To provide insight into the origins of the small changes to the partial ionization of Tyr16 ($I_{tot}$) upon replacement of tyrosine residues by 3-chlorotyrosines, we decompose $I_{tot}$ as
\begin{equation}
I_{tot}=\int{P(R)I(R)dR},
\end{equation}
where $P(R)$ represents the probability distribution of the donor-acceptor distance, $R$, and $I(R)$ is the partial ionization of Tyr16 at a given $R$. This allows us to separate the effects on partial ionization arising directly from the change in the distance between the hydrogen donor and acceptor, $R$, upon perturbation  from those arising from the change in the ionization at a given distance. 
As shown in Fig. \ref{fig:triad_decomp}a, chlorinating Tyr16 shortens the average distance between Tyr16 and Tyr57, $R_{16/57}$, by about 0.03 \AA~compared to the unlabeled protein. Chlorinating Tyr57 has almost no impact and both are consistent with the previous experimental observation that $R_{16/57}$ is unaffected by chlorination within the instrument resolution of $\sim$0.1 \AA\cite{Wu2015}. Due to the minor change in $R_{16/57}$ in Cl-Tyr57 KSI$^{D40N}$, the ionization transferred from Tyr57 to Tyr16 is similarly small (1\%) compared to the unlabeled protein. 

Fig. \ref{fig:triad_decomp}b shows that the ionization depends strongly on the donor-acceptor distance ($\sim4\%$ per 0.1 \AA). Chlorination, which alters the relative proton affinities, is seen to make the largest change at the shortest $R_{16/57}$ values (making a change of $\sim2\%$ at 2.35 \AA) while making almost no difference at 2.8 \AA. From this one can calculate that of the total ionization change of 4\% upon chlorination of Tyr16, 1.5\% comes from the change in $P(R_{16/57})$ and 2.5\% from the change in the slope of $I(R_{16/57})$. Therefore, both the short $R$ and the matching of the proton affinity of the residues in the active-site hydrogen bond network of KSI$^{D40N}$ are almost equally crucial factors in determining the residues' change in partial ionization in the presence of perturbations. The invariance of $I_{tot}$ comes from the fact that chlorination leads to a small change (less than 20\% at any R) in both $P (R_{16/57})$ and $I (R_{16/57})$.

\subsubsection{Change in $pK_a$ isotope effect in response to mutations}
The excess isotope effect, $\Delta\Delta pK_a$, defined as the difference between the H/D $pK_a$ isotope effect in KSI$^{D40N}$ and that in tyrosine solution, provides another experimental measure of the change in the network upon chlorination. We have previously shown that the KSI$^{D40N}$ hydrogen bond network results in a large excess isotope effect for the triad ($\Delta\Delta pK_a=0.50\pm 0.03$) which arises from the close match between the zero-point energy in the O--H bond and the energetic barrier needed to move  a proton across a hydrogen bond from a donor to the acceptor.\cite{Wang2014} If the nuclei were to behave classically, the $pK_a$ would have no dependence on the mass of the protons. Hence $\Delta\Delta pK_a$ arises purely from the quantum mechanical nature of the nuclei, which can be treated using path integral simulations.

\begin{figure}[ht]
\centering
\includegraphics[width=0.75\columnwidth]{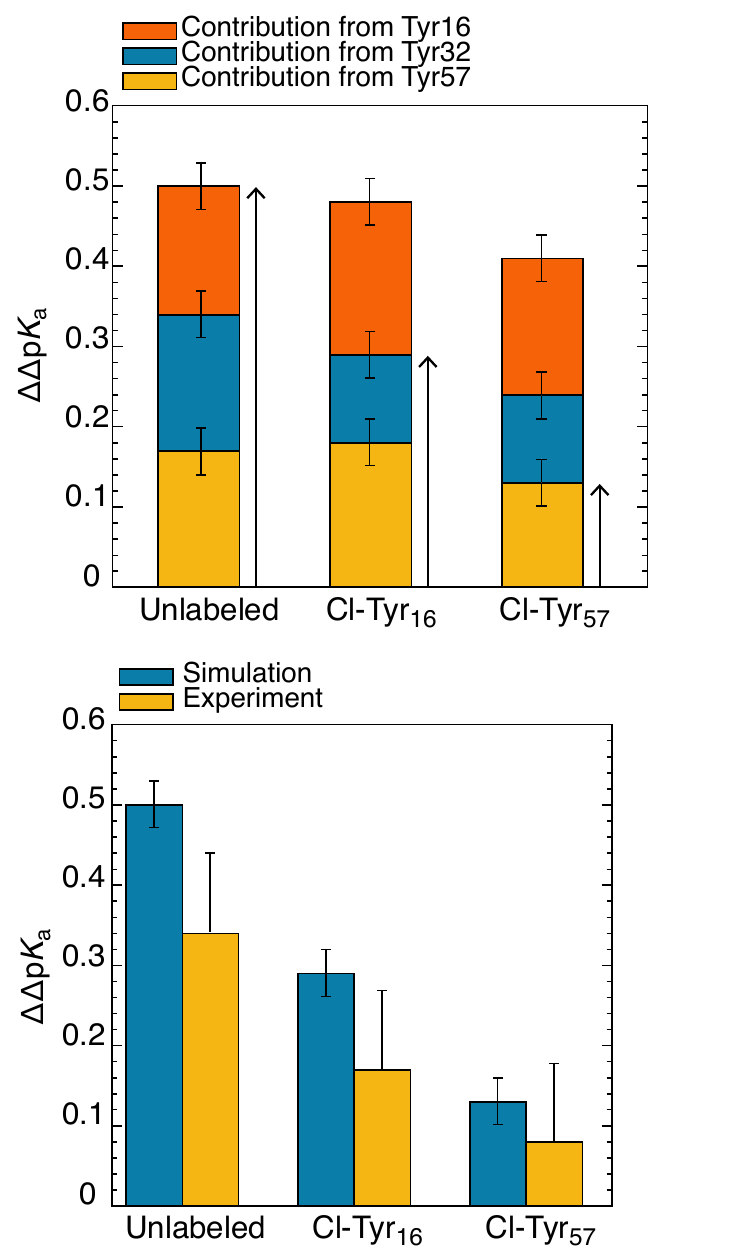}
\caption{$\Delta\Delta pK_a$ for unlabeled, Cl-Tyr16, and Cl-Tyr57 KSI$^{D40N}$. (a) Contributions of Tyr16, Tyr32 and Tyr57 to the overall $\Delta\Delta pK_a$. The arrows show the
residues measured in experiment. (b) Comparison of $\Delta\Delta pK_a$ obtained from AI-PIMD simulations and UV-Vis experiments.\cite{Wu2015} The theoretical and experimental
error bars are 0.03 and 0.10, respectively.}
\label{fig:isotope}
\end{figure}

In Cl-Tyr16 KSI$^{D40N}$, one might expect residue 16 to be more likely to donate its proton due to its lower $pK_a$, which would decrease the proton transfer barrier and enhance the quantum delocalization of protons in the biological hydrogen bond network. This would lead to an increase in $\Delta\Delta pK_a$ as compared to the unlabeled protein. Similarly, one might expect less proton sharing and a much smaller $\Delta\Delta pK_a$ in Cl-Tyr57 KSI$^{D40N}$, since the deprotonated residue 57 can be stabilized by the electron-withdrawing chlorine rather than by hydrogen bonding. However, as shown in Fig. \ref{fig:isotope}a, our AI-PIMD simulations predict that the $\Delta\Delta pK_a$ values are very similar for the unlabeled, Cl-Tyr16 and Cl-Tyr57 KSI$^{D40N}$. Compared to the unlabeled protein, chlorinating Tyr16 has no impact on $\Delta\Delta pK_a$ within the error bar ($\pm$ 0.03), while chlorinating Tyr57 only slightly decreases the isotope effect.

To uncover the origins of the small changes in $\Delta\Delta pK_a$, we again evaluate the impact of the donor acceptor distance $R$ and matching of proton affinity in the triad by decomposing $\Delta\Delta pK_a$ as
\begin{equation}
\Delta\Delta pK_a=\int{P(R)A(R)dR},
\end{equation}
where $A(R)$ is the $\Delta\Delta pK_a$ value obtained at donor-acceptor distance $R$. KSI$^{D40N}$ and its two chlorine-labeled mutants share the same $A (R_{16/57})$ profile within the error bar ($\pm$ 0.05), which suggests that it is the O--O distances, rather than the difference in the residues' proton affinity, that dominate $\Delta\Delta pK_a$ in the hydrogen bonded triad. $A (R_{16/57})$ for the unlabeled protein is shown in Fig. \ref{fig:triad_decomp}c. Thus, unlike the partial ionization, the small change in $\Delta\Delta pK_a$ is nearly entirely due to the largely unchanged distance distribution $P(R_{16/57})$ upon chlorination.

As shown in Fig. \ref{fig:triad_decomp}c, the isotope effect $A(R_{16/57})$ drops by 55\% when $R_{16/57}$ moves from 2.45 \AA~to 2.75 \AA. To understand this trend, we decompose $A(R_{16/57})$ into three orthogonal internal directions -- one along the O--H covalent bond (O--H), one in the plane of the O--H--O hydrogen bond and one perpendicular to the plane of the hydrogen bond. The O--H direction gives a large positive contribution to $\Delta\Delta pK_a$ since the larger ZPE of the O--H stretch motion allows the H atom to be more quantum mechanically delocalized compared to D. This strengthens the hydrogen bond in KSI$^{D40N}$ compared to tyrosine in solution. To illustrate this point, as shown in Fig. \ref{fig:triad_decomp}c, if only the O--H  contribution were present, $\Delta\Delta pK_a$ would be 0.85 for $R$ = 2.65 \AA. However, ZPE in the two directions orthogonal to the O--H direction distorts the hydrogen bond and hence has the opposite effect. These orthogonal contributions reduce $\Delta\Delta pK_a$ by 0.46, resulting in the observed value of 0.39 at that distance. Fig. \ref{fig:triad_decomp}c shows that for every $R_{16/57}$ sampled in AI-PIMD simulations, a large amount of these two effects cancel. This observation demonstrates the established principle of competing quantum effects in hydrogen bonds.\cite{habe+09jcp,li+11pnas,mark-bern12pnas,roma+13jpcl,McKenzie2014,Wang2014,Ceriotti2016,Fang2016}
The amount of cancellation increases from 55\% to 73\% as $R_{16/57}$ moves from 2.45 \AA~to 2.75 \AA, giving rise to the large drop in $A(R_{16/57})$. As shown in Fig. \ref{fig:triad_decomp}c, this increase in the cancellation all arise from the fall in the O--H contribution, with the orthogonal contribution staying roughly constant as $R$ increases. This reflects the fact that small changes in $R$ can dramatically modulate the barrier to delocalize the proton along the O--H direction. The observed $A(R_{16/57})$ is positive for all relevant $R_{16/57}$ and thus contributions along the O--H direction dominate. This is in contrast to the situation observed in liquid water where $R$ $\approx$ 2.8 \AA, leading to the two competing quantum effects almost perfectly cancelling each other.\cite{li+11pnas,mark-bern12pnas} In KSI$^{D40N}$, NQEs act to strength the active-site hydrogen bond network because the overall protein fold keeps the tyrosine residues in such close proximity that the strengthening effect arising from the O--H contribution dominates.

These results allow us to suggest an explanation for two seemingly contradictory observations made in recent experiments: the unlabeled and Cl-Tyr16 KSI$^{D40N}$ were observed to have a large difference in $\Delta\Delta pK_a$, in contradiction to the small changes observed in their partial ionization values.\cite{Wu2015} Our results suggest that the UV-Vis experiments used to extract $\Delta\Delta pK_a$ were complicated by the overlapping spectral signatures of tyrosine and chlorotyrosine. Specifically, in the unlabeled protein, Tyr16, Tyr32 and Tyr57 contribute to a single absorption band in the UV-Vis spectrum and thus the experimentally measured $\Delta\Delta pK_a$ probes the triad as a whole, not the individual residues. However, when a chlorine-substituted tyrosine is incorporated, the band is split into two absorption peaks arising from tyrosinate and 3-chlorotyrosinate, which have significant spectral overlap.\cite{Wu2015} As a result, in UV-Vis measurements of ionization of Cl-Tyr16 KSI$^{D40N}$, only the contributions of ionized Tyr32 and Tyr57 are detected, while in UV-Vis measurements of Cl-Tyr57 KSI$^{D40N}$, only the contribution of ionized Cl-Tyr57 is detected, as shown with arrows in Fig. \ref{fig:isotope}a. In Fig. \ref{fig:isotope}b we show that when we isolate these contributions from our simulations, the resulting $\Delta\Delta pK_a$ are in agreement with experiments within the error bars. 

\subsection{Electric fields in the extended hydrogen bond network}

\begin{figure}[ht]
\centering
\includegraphics[width=0.75\columnwidth]{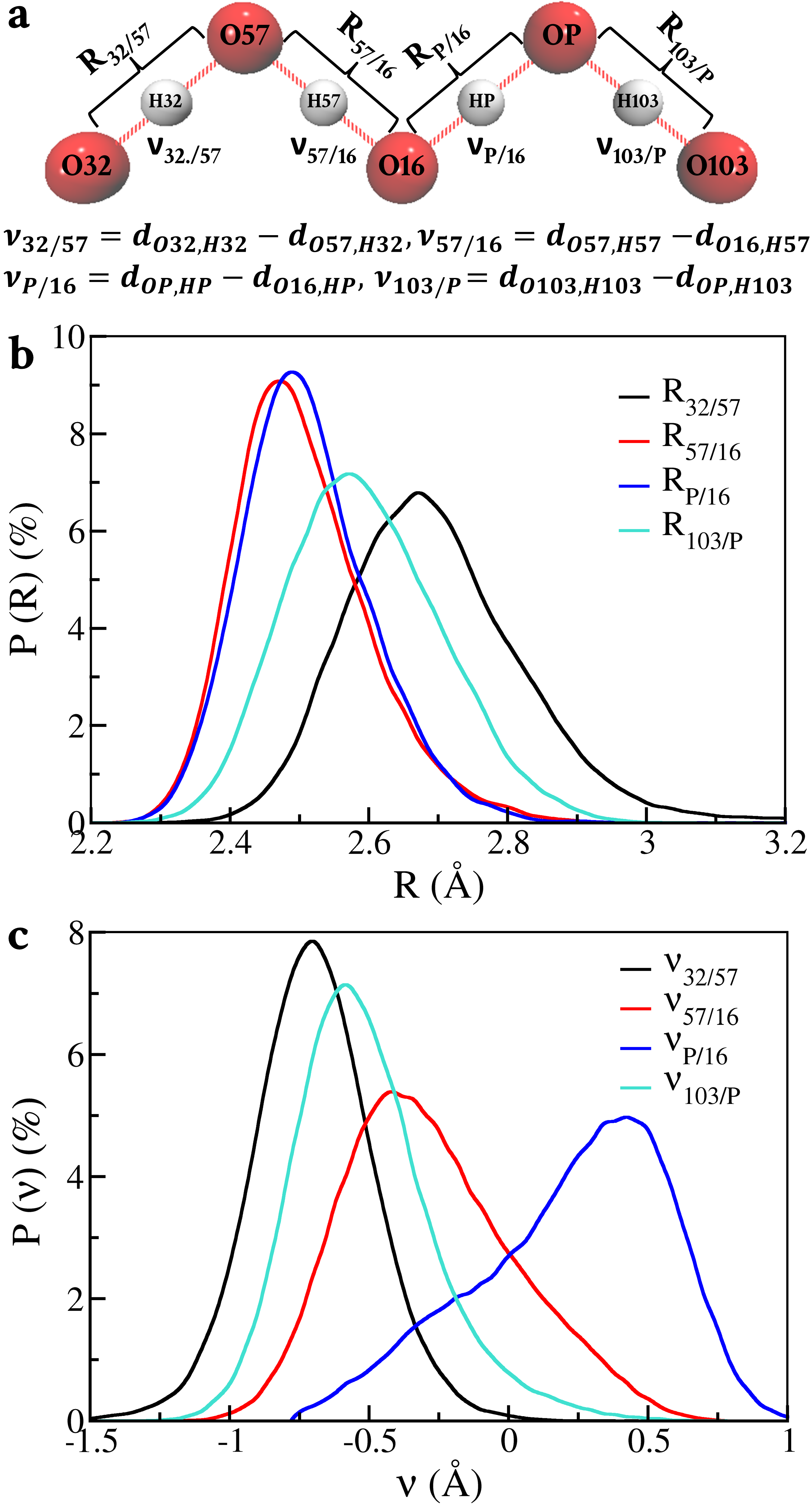}
\caption{Schematic depiction of the extended hydrogen bond network (a) and the distributions of the O--O distances R (b) and the proton transfer coordinate $\nu$ (c).}
\label{fig:Ext}
\end{figure}

Upon docking an intermediate analog, phenol, an extended hydrogen bond network is formed which incorporates phenol and residue Asp103 (Fig. \ref{fig:triad}b). As shown in Fig. \ref{fig:Ext}b, the average distances between the five oxygen atoms in the network from our AI-PIMD simulations are all below 2.7 \AA. In particular, the network is composed of three triads where the central triad is composed of Tyr57, Tyr16 and phenol. The distances between these residues, $R_{57/16}$ and $R_{P/16}$, both fluctuate around an average distance of $\sim$2.5 \AA. The extended hydrogen bond network has shorter O--O distances compared to the tyrosine triad in the apo protein, which allow the protons, in particular H57 and HP, to sample a wide range of distances along the $\nu$ coordinates ($\sim$1.8 \AA), as shown in Fig. \ref{fig:Ext}c.

Due to the delocalization of the protons, the hydrogen bonding partners in the extended network are partially ionized. As Tyr57, Tyr16 and phenol are all at the center of their corresponding triad structures, they are 22\%, 37\% and 21\% ionized, respectively, compared with the experimental values of $\sim$39\%, $\sim$38\% and $\sim$23\%.\cite{Sigala2013} Residues Tyr57 and Tyr16 thus share the ionization of phenol, stabilizing the negatively charged intermediate analog.\cite{Wang2014} 

The ability of KSI to stabilize the charged dienolate intermediate more than its uncharged ground state has been suggested to be connected to the electric fields in the enzyme's active site.\cite{Fried2014}
Recent experiments used Stark effect spectroscopy to show that wild-type KSI exerts a large electric field of -144 $\pm$ 6 MV/cm on the C=O bond of a bound inhibitor 19-nortestosterone (19-NT), while KSI$^{D40N}$ generates a field of -135 $\pm$ 4 MV/cm.\cite{Fried2014} In addition, KSI mutants with the active-site tyrosine residues replaced by 3-chlorotyrosine were observed to retain the large electric field (between -119 and -131 MV/cm depending on the mutation site),\cite{Wu2016} which is consistent with the relatively unperturbed structural arrangement of the active-site residues and the modest change in tyrosine's O--H dipole following chlorination. It further suggests that the large field an inhibitor experiences in this environment arises from the geometric arrangement of the active site. Our QM/MM AI-PIMD simulations give a total electric field experienced by the phenol's C--O bond, $E_{CO}$, of -152 $\pm$ 2 MV/cm, which much more closely agrees with experiment (-135 MV/cm) than previous simulations (-57.6 MV/cm and -60.4 MV/cm).\cite{Fried2014,Bhowmick2017} In both our and the previous simulations using polarizable force fields,\cite{Bhowmick2017} the contribution from the protein scaffold is a small part of the total electric field (-2 MV/cm and -16 MV/cm respectively), but in our case the contribution from the active site is 3.7 fold larger, bringing it into closer agreement with the observed electric field.

\begin{figure}[h]
\centering
\includegraphics[width=0.8\columnwidth]{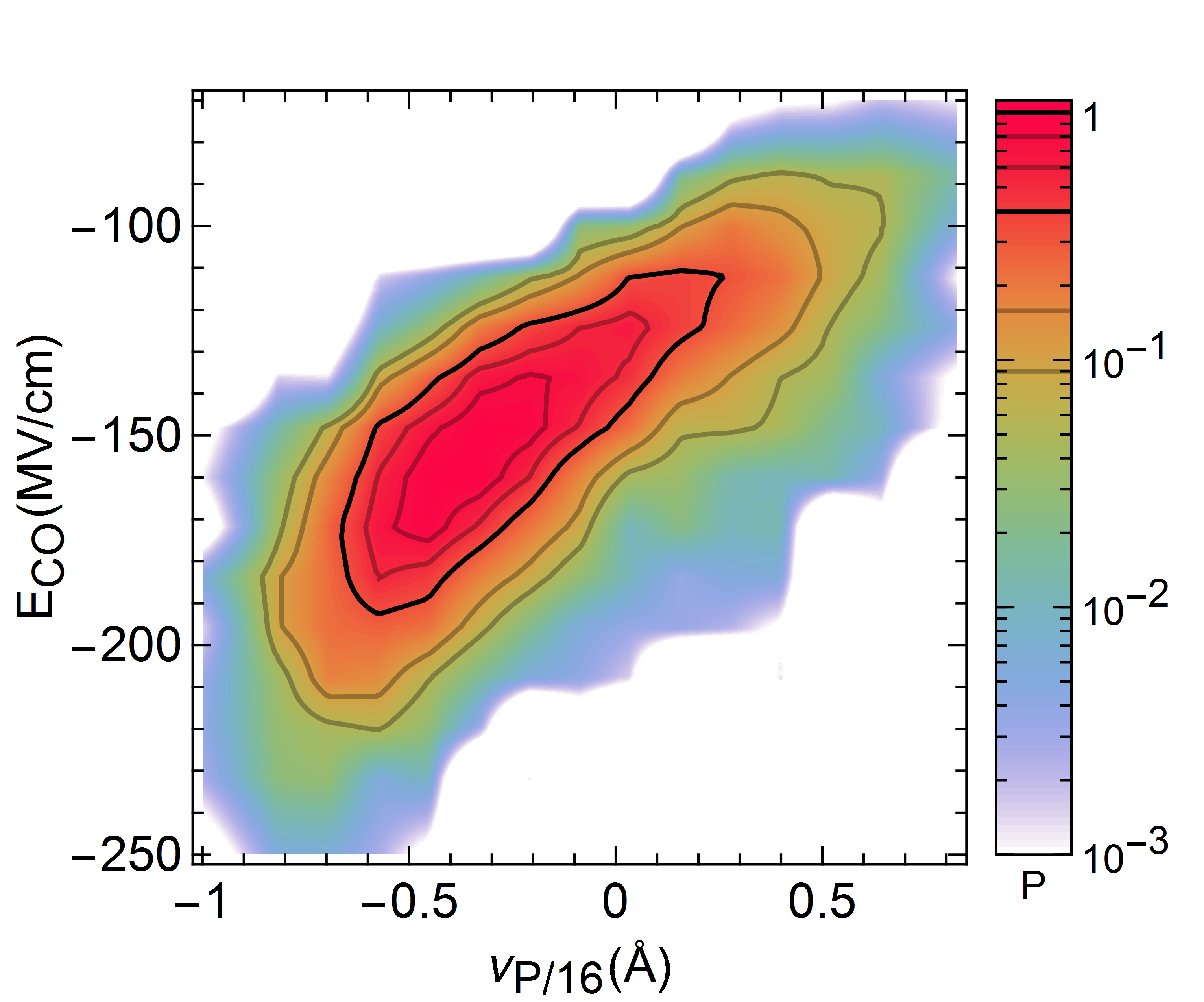}
\caption{Joint probability of $\nu_{P/16}$ and $E_{CO}$ in KSI$^{40N}$ with a phenol bound, as obtained from AI-PIMD simulations and normalized by the maximum probability.}
\label{fig:field}
\end{figure} 

Where does the large electric field come from? By comparing $E_{CO}$ in the presence and absence of the protein environment, we find that the fields exerted by residues in the active-site hydrogen bond network (Fig. \ref{fig:triad}b) account for 98\% of the overall electric field. Considering the large degree of proton sharing between Tyr16 and phenol in this network, we plot $E_{CO}$ as a function of the position of H16, as captured by $\nu_{P/16}=d_{OP,H16}-d_{O16,H16}$. As shown in Fig. \ref{fig:field}, $E_{CO}$ fluctuates as the proton and thus the electron move along the hydrogen bond, and increases dramatically in magnitude when the proton moves towards the intermediate analog (i.e, when $\nu_{P/16}$ becomes more negative). The large average $E_{CO}$ (-152 MV/cm) thus results from a short average $\nu_{P/16}$ of 0.1 \AA~in the AI-PIMD simulations (Fig. \ref{fig:Ext}c), which is close to the value of zero expected for a proton which sits on average exactly equidistant from both residues. Our calculated $E_{CO}$ is slightly larger than the experimentally measured fields.\cite{Fried2014} One likely reason for this is that phenol in the simulations is an intermediate analog, which should allow for greater proton sharing with Tyr16 than 19-NT in the experiment, which is a substrate analog that possesses a C=O group (in place of C--O). Therefore, the calculated average $E_{CO}$ from the simulations can be viewed as an upper limit to the electric field that the C=O in 19-NT would experience. Importantly, this calculation is consistent with the suggestion that KSI's substrate undergoes small geometric changes during the catalytic cycle, and hence the electric field experienced by the C=O bond in the ground state is similar to that experienced by the C--O bond in the transition state/intermediate state.\cite{Fried2017}

Since the active-site hydrogen bond network gives rise to most of $E_{CO}$, we can now further decompose the electric field into contributions from different residues. We performed calculations on KSI$^{D40N}$ in the absence and presence of individual active-site residues, and found that Tyr16 and Asp103 dominate $E_{CO}$ as they form direct hydrogen bonds to the intermediate analog. The values of $E_{CO}$ from Tyr16 and Asp103 are -97 $\pm$ 2 MV/cm and -47 $\pm$ 1 MV/cm, respectively. These decomposition results are in good agreement with the experimental values of -84 $\pm$ 7 MV/cm and -52 $\pm$ 7 MV/cm, which were obtained by making the Tyr16Phe and Asp103Asn mutations.\cite{Fried2014,Footnote}

\begin{figure}[h]
\centering
\includegraphics[width=0.9\columnwidth]{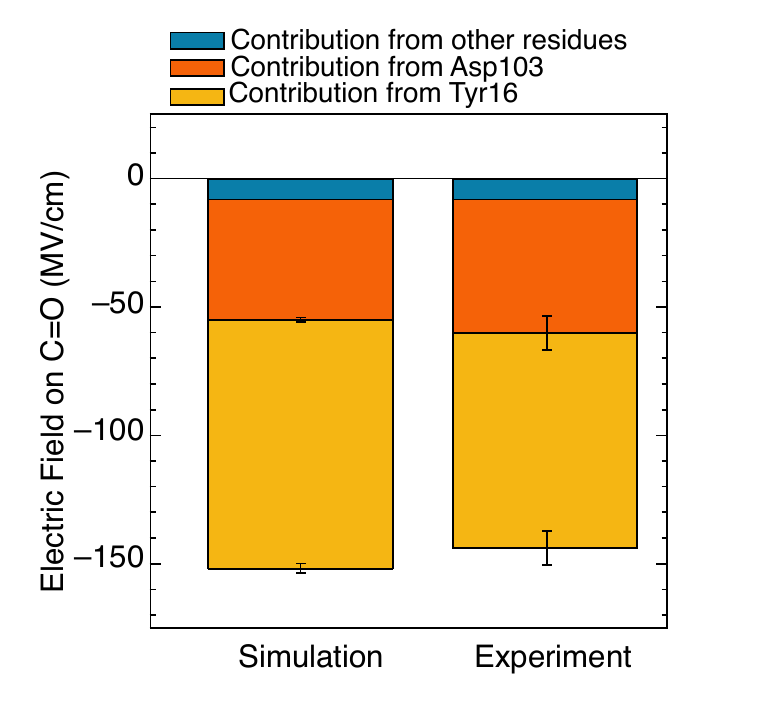}
\caption{Decomposition of the average $E_{CO}$ on the C-O bond of phenol (from AI-PIMD simulations) and on the C=O bond of 19-NT (from experiments\cite{Fried2014}).}
\label{fig:field_decomp}
\end{figure} 

However, it is essential to note that, while the protein environment is not a major contributor to the net electric fields, it plays a crucial role of maintaining the short O--O distances and orientations in the extended hydrogen bond network, which would otherwise not be stable. This positioning is essential: for example, in AI-PIMD simulations of KSI$^{D40N}$ with the bound phenol, the average $R_{P/16}$ is 2.5 \AA, whereas in previous classical molecular dynamics simulations, the steep repulsive walls in the force field push the O--O distances in the active-site hydrogen bond network to more than 4 \AA, which lead to them to form weaker hydrogen bonds and hence to under-predict the $E_{CO}$ value by a factor of $\sim$2.7.\cite{Fried2014}

Hence, the rigid protein environment which stabilizes the heavy atoms’ geometry permits the protons to flexibly move within the network (Fig. \ref{fig:Ext}). The flexibility of the protons leads simultaneously to large but also quite heterogeneous electric fields (Fig. \ref{fig:field}). Experiments have shown that the carbonyl stretch of 19-NT exhibits a very narrow IR linewidth;\cite{Fried2014} however, these findings are consistent with the field fluctuations shown in Fig.~\ref{fig:field} being associated primarily with movement of the protons, not the heavy atoms. Although AI-PIMD simulations do not provide dynamical information, we expect the proton movements occur on an ultrafast timescale, and hence would be averaged out on the IR timescale.

\section{Conclusions}
Here we have shown that electronic and nuclear quantum fluctuations in the active-site tyrosine triad in KSI$^{D40N}$ give rise to small changes in the residues' partial ionizations and $\Delta\Delta pK_a$ in response to perturbations such as unnatural amino acid mutagenesis, by shifting the proton positions and redistributing the electron density to buffer the perturbation effects. In the extended hydrogen bond network upon binding an intermediate analog, quantum fluctuations also result in a large electric field in the enzyme's active site. We have demonstrated how this flexibility and the large electric fields arise from the fact that the short O--O distances lower the potential energy barrier for transferring protons along the hydrogen bond network and that NQEs further flood the potential energy well, thus allowing the electrons and protons to move over a wide range of distances. This work further highlights how AI-PIMD simulations of biological systems can be powerful tools that can reproduce and explain experimental observables such as isotope effects and electric fields.\cite{Wang2014a,Rossi2015,Shi2015,Wang2016,Fang2016,Pinotsi2016}

Finally, it is worth commenting on how the flexibility of the hydrogen bond network might impact the enzyme's function. For example, our observation of the small change in the ionization of the residues upon chlorination is consistent with the recent experimental observation that replacement of Tyr16 with a chloro- or fluoro-tyrosine leads to very little change to KSI's catalytic activity.\cite{Natarajan2014} In addition, the ability to form an extended hydrogen bond network with enhanced proton flexibility, rather than a single hydrogen bond, may also be of importance. In particular, a single Tyr57Phe or Asp103Leu mutation, both of which retain an extended network where the substrate can dock, lead to modest 6 or 100 fold decrease in KSI's catalytic activity, respectively.\cite{Kim2000,Choi2001} However, the combined Tyr57Phe/Asp103Leu mutation, which leaves Tyr16 as the only hydrogen bond donor to the substrate with no extended network and might be expected to yield a 600 fold decrease if the effects of these mutations were independent, reduces the activity by over 15,800 fold.\cite{Kim2000} This suggests that Tyr57 can partially compensate for the impact of mutating Asp103 since the substrate can still be incorporated in an extended hydrogen bond network, which possesses electronic and nuclear flexibility. This robustness of the network thus might protect the enzyme from losing its function unless multiple mutations occur concurrently.  

\begin{acknowledgments}
We thank Yufan Wu and Professor Steven Boxer for discussions of the details of their experiments. This material is based upon work supported by the National Science Foundation under Grant No. CHE-1652960. T.E.M also acknowledges support from a Cottrell Scholarship from the Research Corporation for Science Advancement and the Camille Dreyfus Teacher-Scholar Awards Program. L.W. acknowledges a postdoctoral fellowship from the Stanford Center for Molecular Analysis and Design, and the HPC facilities from the Office of Advanced Research Computing at Rutgers University. S.D.F. is supported by a Junior Research Fellowship at King’s College, Cambridge.  This work used the Extreme Science and Engineering Discovery Environment, which is supported by NSF Grant ACI-1053575 (Project TG-CHE140013). This work used the XStream computational resource, supported by the National Science Foundation Major Research Instrumentation program (ACI-1429830). 
\end{acknowledgments}

\providecommand{\latin}[1]{#1}
\providecommand*\mcitethebibliography{\thebibliography}
\csname @ifundefined\endcsname{endmcitethebibliography}
  {\let\endmcitethebibliography\endthebibliography}{}

\clearpage
\onecolumngrid

\appendix

\setcounter{equation}{0}
\setcounter{figure}{0}
\makeatletter 
\renewcommand{\thefigure}{S\@arabic\c@figure}
\makeatother
\renewcommand{\theequation}{S\arabic{equation}}

\section*{Supporting Information}
\subsection{AIMD and AI-PIMD simulations}
In the AIMD and AI-PIMD simulations, the QM regions for KSI$^{D40N}$ and chlorinated KSI$^{D40N}$ contain the p-methylene phenol side chains
 of residues 16, 32 and 57. For KSI$^{D40N}$ with phenol
bound, the QM regions also include the side chain of Asp103 and phenol.
The MM regions contains the rest of the protein, sodium counter-ions and solvating water molecules. For tyrosine in aqueous solution, the QM region contains the p-methylene phenol side chain of tyrosine and 41 surrouding water molecules. The MM region contains 1679 water molecules.   

Each simulation was performed in the NVT ensemble at 300 K for a total lengths of 20 ps with a time step of 0.5 fs.
In the AIMD and AI-PIMD simulations, the electronic structure in the QM region was evaluated at each step
at the B3LYP level\cite{beck93jcp} with the D3 dispersion corrections,\cite{grim+10jcp} combined with the 6-31G* 
basis set.
In addition, in the AI-PIMD simulations each particle was represented by six path integral beads to account for nuclear quantum effects,
using the path integral generalized Langevin equation approach.\cite{ceri-mano12prl} The MM region was modeled
using classical dynamics with the AMBER03 force field\cite{Duan2003} and the TIP3P water model.\cite{jorg+83jcp} The simulations were performed with an in-house code. MPI interfaces with the TeraChem\cite{Ufimtsev2009,Isborn2012} and LAMMPS\cite{plim95jcp} packages were used to evaluate the interactions
within the QM and MM regions, respectively, and account for the electrostatic and van der Waals interactions between the QM and MM regions. Bonds across the QM-MM boundary were
capped with hydrogen atoms, with the linking atoms included in the QM region and having no interactions with the MM atoms. 

\subsection{Linear interpolation in partial ionization calculations}
To justify the linear interpolation scheme we used to calculate the residues' partial ionization values, we stretched the O16--H16 bond along the O16--O57 vector and calculated the corresponding $^{13}$C NMR chemical shifts. This calculation was performed on the crystal structure of KSI (PDB ID 1OGX).\cite{Ha2000} The QM region contains the tyrosine triad, while all the other protein atoms are treated as point charges according to the AMBER03 force field.\cite{Duan2003} The B3LYP functional\cite{beck93jcp} and 6-31G* basis set were used, as implemented in the Gaussian 09 program.\cite{g09} The resulting $^{13}$C NMR chemical shifts as a function of the O16--H16 length is shown in Fig. S2.

\subsection{$\Delta\Delta pK_a$ calculation}
Calculations of the excess isotope effect $\Delta\Delta pK_a$ followed the methods described in Ref 1. Briefly, $\Delta\Delta pK_a=\Delta pK_a^{KSI}-\Delta pK_a^{Sol}$, where $\Delta pK_a^{KSI}$ is the $pK_a$ isotope effect in KSI$^{D40N}$ upon substituting proton (H) by deuterium (D) and $\Delta pK_a^{Sol}$ is that of tyrosine in aqueous solution. $\Delta\Delta pK_a$ was calculated using the thermodynamic cycles shown in Fig. S3. Here KSIH and KISD are KSI$^{D40N}$ with Tyr57 protonated in H$_2$O or D$_2$O, respectively. KSI$^-$ and KSID$^-$ are KSI$^{D40N}$ with Tyr57 deprotonated in H$_2$O and D$_2$O, respectively. Similarly, TyrH and TyrD are neutral tyrosine in H$_2$O and D$_2$O, respectively, and Tyr$^-$ is the deprotonated tyrosine in aqueous solution. If we label the free energy changes shown in Fig. S3 as $\Delta A$, then the isotope effects are
\begin{equation*}
    \Delta pK_a^{KSI}=\frac{\Delta A_2-\Delta A_1}{2.303RT}=\frac{\Delta A_{KSIH}-\Delta A_{KSI^-}-\Delta A_H}{2.303RT},
\end{equation*}
\begin{equation*}
    \Delta pK_a^{Sol}=\frac{\Delta A_4-\Delta A_3}{2.303RT}=\frac{\Delta A_{TyrH}-\Delta A_H}{2.303RT}.
\end{equation*}
Here R is the gas constant and T is the temperature. Thus the excess isotope effect is 
\begin{equation}
    \Delta\Delta pK_a=\Delta pK_a^{KSI}-\Delta pK_a^{Sol}=\frac{\Delta A_{KSIH}-\Delta A_{KSI^-}-\Delta A_{TyrH}}{2.303RT}.
\end{equation}
The free energy terms were calculated from the quantum kinetic energies of the hydrogen isotopes,\cite{ceri-mark13jcp,Wang2014a}
\begin{equation}
    \Delta A_i=-\int_{m_D}^{m_H}d\mu\frac{\langle K_i(\mu)\rangle}{\mu}.
\end{equation}
where $K_i(\mu)$ is the quantum kinetic energy of a hydrogen isotope of mass $\mu$. $K_i(H)$ was obtained from AI-PIMD simulations of KSI$^{D40N}$ and tyrosine in solution using the centroid virial estimator.\cite{herm-bern82jcp,cao-bern89jcp}, while $K_i(\mu)$ was calculated using the thermodynamic free-energy perturbation (TD-FEP) path integral estimator.\cite{ceri-mark13jcp} 

If one uses the kinetic energies of protons H16, H32 or H57 in Eq. 2, one can calculate the contributions of Tyr16, Tyr32 and Tyr57, respectively, to $\Delta\Delta pK_a$. The contributions of Tyr16 and Tyr32 were obtained by comparing the free energy differences of the protons in the neutral and ionized KSI$^{D40N}$. For example, for Tyr16
\begin{equation}
  \Delta\Delta pK_a^{Tyr16}=\frac{\Delta A_{KSIH}^{Tyr16}-\Delta A_{KSI^-}^{Tyr16}}{2.303RT},
\end{equation}
where $\Delta A_{KSIH}^{Tyr16}$ and $\Delta A_{KSI^-}^{Tyr16}$ are the free energy changes of H16 in the neutral and ionized protein. The contribution of H57 was calculated by comparing H57 and tyrosine in solution,
\begin{equation}
   \Delta\Delta pK_a^{Tyr57}=\frac{\Delta A_{KSIH}^{Tyr57}-\Delta A_{TyrH}}{2.303RT}.
\end{equation}

One can further decompose $\Delta\Delta pK_a$ in three directions. Here direction 1 is along the O--H bond, direction 2 is in the hydrogen bonded O--H--O plane and orthogonal to direction 1, and direction 3 is perpendicular to the hydrogen bond plane. By decomposing the quantum kinetic energies in Eq. 2 in these three orthogonal directions, we can illustrate the competiting quantum effects. The decomposition results are shown in Fig. 3 and 4 in the main text.  

\section*{SI figures}

\begin{figure}[h!tbp]
\centering
\includegraphics[height=4.2cm]{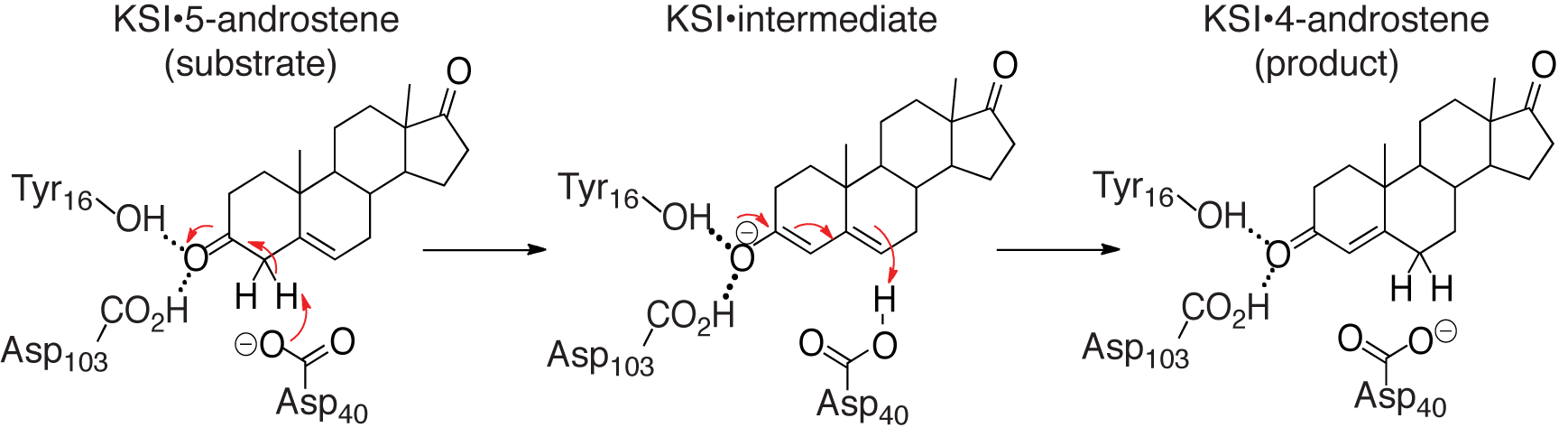}
\caption{Catalytic mechanism of KSI.}
\end{figure} 

\begin{figure}[h!tbp]
\centering
\includegraphics[height=7cm]{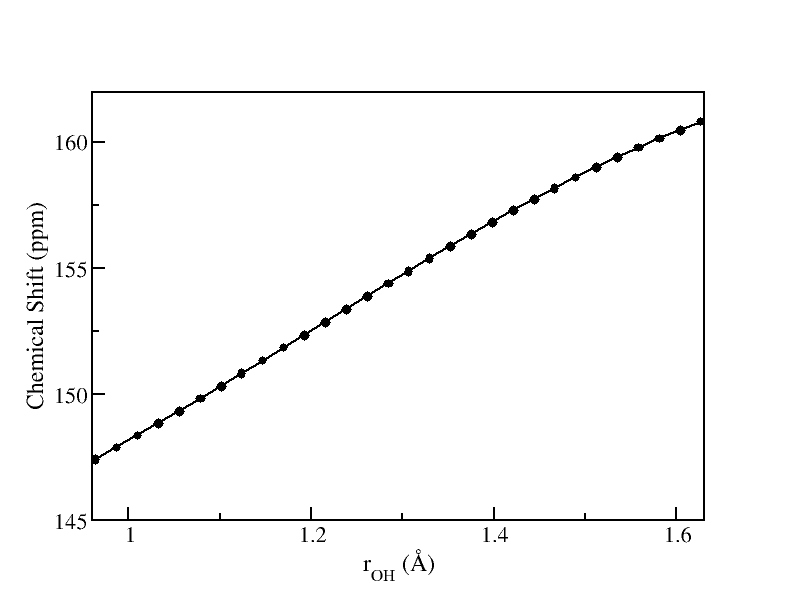}
\caption{$^{13}$C NMR chemical shifts when the proton H16 is moved between Tyr16 and Tyr57, calculated from the crystal structure of KSI$^{D40N}$ (PDB ID 1OGX).\cite{Ha2000}}
\end{figure}

\begin{figure}[h!tbp]
\centering
\includegraphics[height=4cm]{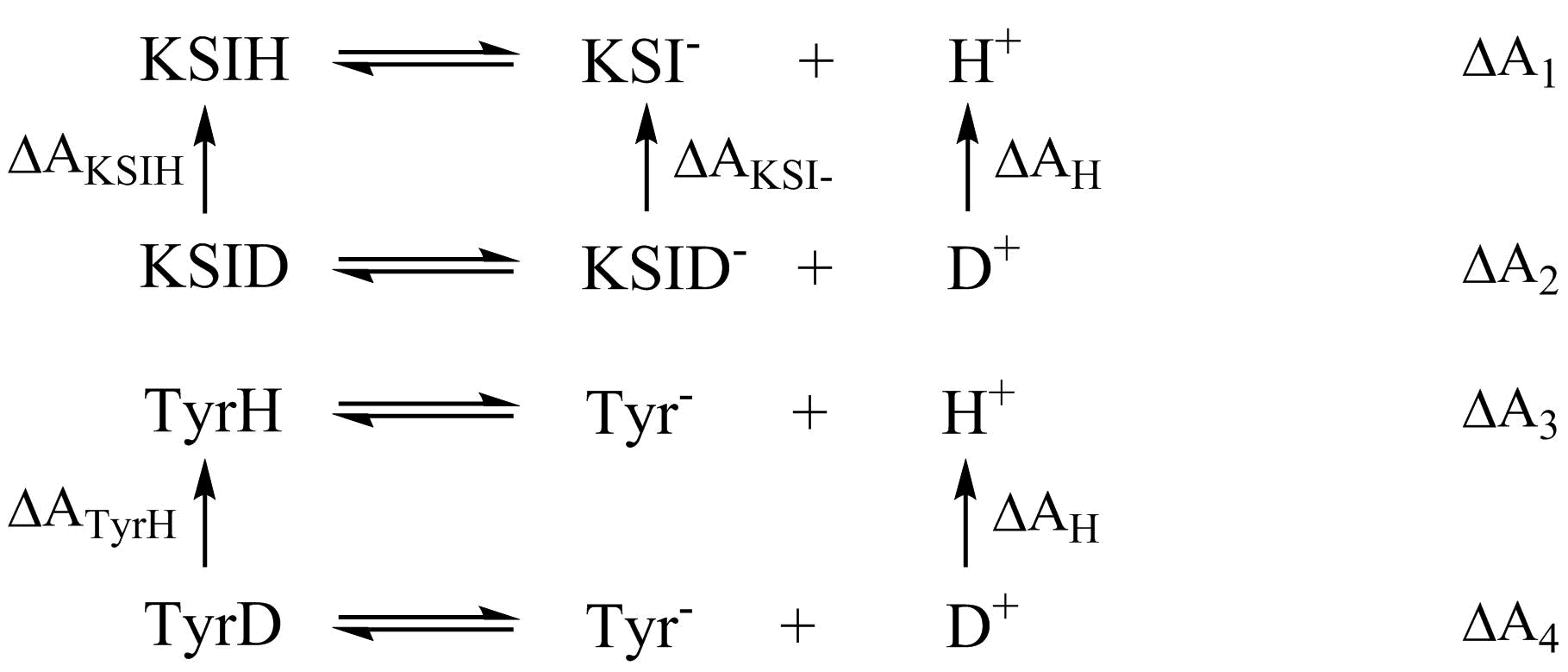}
\caption{Thermodynamic cycles used for $\Delta\Delta pK_a$ calculations.}
\end{figure} 

\providecommand{\latin}[1]{#1}
\providecommand*\mcitethebibliography{\thebibliography}
\csname @ifundefined\endcsname{endmcitethebibliography}
  {\let\endmcitethebibliography\endthebibliography}{}


\begin{mcitethebibliography}{71}
\providecommand*\natexlab[1]{#1}
\providecommand*\mciteSetBstSublistMode[1]{}
\providecommand*\mciteSetBstMaxWidthForm[2]{}
\providecommand*\mciteBstWouldAddEndPuncttrue
  {\def\EndOfBibitem{\unskip.}}
\providecommand*\mciteBstWouldAddEndPunctfalse
  {\let\EndOfBibitem\relax}
\providecommand*\mciteSetBstMidEndSepPunct[3]{}
\providecommand*\mciteSetBstSublistLabelBeginEnd[3]{}
\providecommand*\EndOfBibitem{}
\mciteSetBstSublistMode{f}
\mciteSetBstMaxWidthForm{subitem}{(\alph{mcitesubitemcount})}
\mciteSetBstSublistLabelBeginEnd
  {\mcitemaxwidthsubitemform\space}
  {\relax}
  {\relax}

\bibitem[Pauling \latin{et~al.}(1951)Pauling, Corey, and Branson]{Pauling1951}
Pauling,~L.; Corey,~R.~B.; Branson,~H.~R. The structure of proteins: Two
  hydrogen-bonded helical configurations of the polypeptide chain. \emph{Proc.
  Natl. Acad. Sci. U.S.A.} \textbf{1951}, \emph{37}, 205--211\relax
\mciteBstWouldAddEndPuncttrue
\mciteSetBstMidEndSepPunct{\mcitedefaultmidpunct}
{\mcitedefaultendpunct}{\mcitedefaultseppunct}\relax
\EndOfBibitem
\bibitem[Dill(1990)]{Dill1990}
Dill,~K.~A. Dominant forces in protein folding. \emph{Biochemistry}
  \textbf{1990}, \emph{29}, 7133--7155\relax
\mciteBstWouldAddEndPuncttrue
\mciteSetBstMidEndSepPunct{\mcitedefaultmidpunct}
{\mcitedefaultendpunct}{\mcitedefaultseppunct}\relax
\EndOfBibitem
\bibitem[Nelson \latin{et~al.}(2008)Nelson, Lehninger, and Cox]{Nelson2008}
Nelson,~D.; Lehninger,~A.; Cox,~M. \emph{Lehninger Principles of Biochemistry};
  W. H. Freeman, 2008\relax
\mciteBstWouldAddEndPuncttrue
\mciteSetBstMidEndSepPunct{\mcitedefaultmidpunct}
{\mcitedefaultendpunct}{\mcitedefaultseppunct}\relax
\EndOfBibitem
\bibitem[Zhang \latin{et~al.}(2016)Zhang, Rajendram, Weibel, Yethiraj, and
  Cui]{Zhang2016}
Zhang,~L.; Rajendram,~M.; Weibel,~D.~B.; Yethiraj,~A.; Cui,~Q. Ionic Hydrogen
  Bonds and Lipid Packing Defects Determine the Binding Orientation and
  Insertion Depth of RecA on Multicomponent Lipid Bilayers. \emph{J. Phys.
  Chem. B} \textbf{2016}, \emph{120}, 8424--8437\relax
\mciteBstWouldAddEndPuncttrue
\mciteSetBstMidEndSepPunct{\mcitedefaultmidpunct}
{\mcitedefaultendpunct}{\mcitedefaultseppunct}\relax
\EndOfBibitem
\bibitem[Cui(2016)]{Cui2016}
Cui,~Q. Perspective: Quantum mechanical methods in biochemistry and biophysics.
  \emph{J. Chem. Phys.} \textbf{2016}, \emph{145}, 140901\relax
\mciteBstWouldAddEndPuncttrue
\mciteSetBstMidEndSepPunct{\mcitedefaultmidpunct}
{\mcitedefaultendpunct}{\mcitedefaultseppunct}\relax
\EndOfBibitem
\bibitem[Hibbert and Emsley(1990)Hibbert, and Emsley]{Hibbert1990}
Hibbert,~F.; Emsley,~J. Hydrogen Bonding and Chemical Reactivity. \emph{Adv.
  Phys. Org. Chem.} \textbf{1990}, \emph{26}, 255 -- 379\relax
\mciteBstWouldAddEndPuncttrue
\mciteSetBstMidEndSepPunct{\mcitedefaultmidpunct}
{\mcitedefaultendpunct}{\mcitedefaultseppunct}\relax
\EndOfBibitem
\bibitem[Cleland and Kreevoy(1994)Cleland, and Kreevoy]{Cleland1994}
Cleland,~W.; Kreevoy,~M. Low-barrier hydrogen bonds and enzymic catalysis.
  \emph{Science} \textbf{1994}, \emph{264}, 1887--1890\relax
\mciteBstWouldAddEndPuncttrue
\mciteSetBstMidEndSepPunct{\mcitedefaultmidpunct}
{\mcitedefaultendpunct}{\mcitedefaultseppunct}\relax
\EndOfBibitem
\bibitem[Tuckerman \latin{et~al.}(1997)Tuckerman, Marx, Klein, and
  Parrinello]{Tuckerman1997}
Tuckerman,~M.~E.; Marx,~D.; Klein,~M.~L.; Parrinello,~M. On the Quantum Nature
  of the Shared Proton in Hydrogen Bonds. \emph{Science} \textbf{1997},
  \emph{275}, 817--820\relax
\mciteBstWouldAddEndPuncttrue
\mciteSetBstMidEndSepPunct{\mcitedefaultmidpunct}
{\mcitedefaultendpunct}{\mcitedefaultseppunct}\relax
\EndOfBibitem
\bibitem[Perrin \latin{et~al.}(1997)Perrin, , and Nielson]{Perrin1997}
Perrin,~C.~L.; ; Nielson,~J.~B. "Strong" hydrogen bonds in chemistry and
  biology. \emph{Annu. Rev. Phys. Chem.} \textbf{1997}, \emph{48},
  511--544\relax
\mciteBstWouldAddEndPuncttrue
\mciteSetBstMidEndSepPunct{\mcitedefaultmidpunct}
{\mcitedefaultendpunct}{\mcitedefaultseppunct}\relax
\EndOfBibitem
\bibitem[Cleland \latin{et~al.}(1998)Cleland, Frey, and Gerlt]{Cleland1998}
Cleland,~W.~W.; Frey,~P.~A.; Gerlt,~J.~A. The Low Barrier Hydrogen Bond in
  Enzymatic Catalysis. \emph{J. Biol. Chem.} \textbf{1998}, \emph{273},
  25529--25532\relax
\mciteBstWouldAddEndPuncttrue
\mciteSetBstMidEndSepPunct{\mcitedefaultmidpunct}
{\mcitedefaultendpunct}{\mcitedefaultseppunct}\relax
\EndOfBibitem
\bibitem[Marx(2006)]{marx06cpc}
Marx,~D. {Proton transfer 200 years after von Grotthuss: insights from ab
  initio simulations.} \emph{ChemPhysChem} \textbf{2006}, \emph{7},
  1848--70\relax
\mciteBstWouldAddEndPuncttrue
\mciteSetBstMidEndSepPunct{\mcitedefaultmidpunct}
{\mcitedefaultendpunct}{\mcitedefaultseppunct}\relax
\EndOfBibitem
\bibitem[McKenzie(2012)]{mckenzie2012}
McKenzie,~R.~H. {A diabatic state model for donor-hydrogen vibrational
  frequency shifts in hydrogen bonded complexes}. \emph{Chem. Phys. Lett.}
  \textbf{2012}, \emph{535}, 196--200\relax
\mciteBstWouldAddEndPuncttrue
\mciteSetBstMidEndSepPunct{\mcitedefaultmidpunct}
{\mcitedefaultendpunct}{\mcitedefaultseppunct}\relax
\EndOfBibitem
\bibitem[McKenzie \latin{et~al.}(2014)McKenzie, Bekker, Athokpam, and
  Ramesh]{McKenzie2014}
McKenzie,~R.~H.; Bekker,~C.; Athokpam,~B.; Ramesh,~S.~G. Effect of quantum
  nuclear motion on hydrogen bonding. \emph{J. Chem. Phys.} \textbf{2014},
  \emph{140}, 174508\relax
\mciteBstWouldAddEndPuncttrue
\mciteSetBstMidEndSepPunct{\mcitedefaultmidpunct}
{\mcitedefaultendpunct}{\mcitedefaultseppunct}\relax
\EndOfBibitem
\bibitem[Gerlt and Gassman(1993)Gerlt, and Gassman]{Gerlt1993}
Gerlt,~J.~A.; Gassman,~P.~G. An explanation for rapid enzyme-catalyzed proton
  abstraction from carbon acids: importance of late transition states in
  concerted mechanisms. \emph{J. Am. Chem. Soc.} \textbf{1993}, \emph{115},
  11552--11568\relax
\mciteBstWouldAddEndPuncttrue
\mciteSetBstMidEndSepPunct{\mcitedefaultmidpunct}
{\mcitedefaultendpunct}{\mcitedefaultseppunct}\relax
\EndOfBibitem
\bibitem[Yamaguchi \latin{et~al.}(2009)Yamaguchi, Kamikubo, Kurihara, Kuroki,
  Niimura, Shimizu, Yamazaki, and Kataoka]{Yamaguchi2009}
Yamaguchi,~S.; Kamikubo,~H.; Kurihara,~K.; Kuroki,~R.; Niimura,~N.;
  Shimizu,~N.; Yamazaki,~Y.; Kataoka,~M. Low-barrier hydrogen bond in
  photoactive yellow protein. \emph{Proc. Natl. Acad. Sci. U.S.A.}
  \textbf{2009}, \emph{106}, 440--444\relax
\mciteBstWouldAddEndPuncttrue
\mciteSetBstMidEndSepPunct{\mcitedefaultmidpunct}
{\mcitedefaultendpunct}{\mcitedefaultseppunct}\relax
\EndOfBibitem
\bibitem[Pinotsi \latin{et~al.}(2016)Pinotsi, Grisanti, Mahou, Gebauer,
  Kaminski, Hassanali, and Kaminski~Schierle]{Pinotsi2016}
Pinotsi,~D.; Grisanti,~L.; Mahou,~P.; Gebauer,~R.; Kaminski,~C.~F.;
  Hassanali,~A.; Kaminski~Schierle,~G.~S. Proton Transfer and
  Structure-Specific Fluorescence in Hydrogen Bond-Rich Protein Structures.
  \emph{J. Am. Chem. Soc.} \textbf{2016}, \emph{138}, 3046--3057\relax
\mciteBstWouldAddEndPuncttrue
\mciteSetBstMidEndSepPunct{\mcitedefaultmidpunct}
{\mcitedefaultendpunct}{\mcitedefaultseppunct}\relax
\EndOfBibitem
\bibitem[Murgida and Hildebrandt(2001)Murgida, and Hildebrandt]{Murgida2001}
Murgida,~D.~H.; Hildebrandt,~P. Proton-Coupled Electron Transfer of Cytochrome
  c. \emph{J. Am. Chem. Soc.} \textbf{2001}, \emph{123}, 4062--4068\relax
\mciteBstWouldAddEndPuncttrue
\mciteSetBstMidEndSepPunct{\mcitedefaultmidpunct}
{\mcitedefaultendpunct}{\mcitedefaultseppunct}\relax
\EndOfBibitem
\bibitem[Ball(2008)]{Ball2008}
Ball,~P. Water as an Active Constituent in Cell Biology. \emph{Chem. Rev.}
  \textbf{2008}, \emph{108}, 74--108\relax
\mciteBstWouldAddEndPuncttrue
\mciteSetBstMidEndSepPunct{\mcitedefaultmidpunct}
{\mcitedefaultendpunct}{\mcitedefaultseppunct}\relax
\EndOfBibitem
\bibitem[Reece and Nocera(2009)Reece, and Nocera]{Reece2009}
Reece,~S.~Y.; Nocera,~D.~G. Proton-Coupled Electron Transfer in Biology:
  Results from Synergistic Studies in Natural and Model Systems. \emph{Annu.
  Rev. Biochem.} \textbf{2009}, \emph{78}, 673--699\relax
\mciteBstWouldAddEndPuncttrue
\mciteSetBstMidEndSepPunct{\mcitedefaultmidpunct}
{\mcitedefaultendpunct}{\mcitedefaultseppunct}\relax
\EndOfBibitem
\bibitem[Ha \latin{et~al.}(2000)Ha, Kim, Lee, Choi, and Oh]{Ha2000}
Ha,~N.-C.; Kim,~M.-S.; Lee,~W.; Choi,~K.~Y.; Oh,~B.-H. Detection of Large pKa
  Perturbations of an Inhibitor and a Catalytic Group at an Enzyme Active Site,
  a Mechanistic Basis for Catalytic Power of Many Enzymes. \emph{J. Biol.
  Chem.} \textbf{2000}, \emph{275}, 41100--41106\relax
\mciteBstWouldAddEndPuncttrue
\mciteSetBstMidEndSepPunct{\mcitedefaultmidpunct}
{\mcitedefaultendpunct}{\mcitedefaultseppunct}\relax
\EndOfBibitem
\bibitem[Kraut \latin{et~al.}(2006)Kraut, Sigala, Pybus, Liu, Ringe, Petsko,
  and Herschlag]{Kraut2006}
Kraut,~D.~A.; Sigala,~P.~A.; Pybus,~B.; Liu,~C.~W.; Ringe,~D.; Petsko,~G.~A.;
  Herschlag,~D. Testing Electrostatic Complementarity in Enzyme Catalysis:
  Hydrogen Bonding in the Ketosteroid Isomerase Oxyanion Hole. \emph{PLoS
  Biol.} \textbf{2006}, \emph{4}, e99\relax
\mciteBstWouldAddEndPuncttrue
\mciteSetBstMidEndSepPunct{\mcitedefaultmidpunct}
{\mcitedefaultendpunct}{\mcitedefaultseppunct}\relax
\EndOfBibitem
\bibitem[Sigala \latin{et~al.}(2013)Sigala, Fafarman, Schwans, Fried, Fenn,
  Caaveiro, Pybus, Ringe, Petsko, Boxer, and Herschlag]{Sigala2013}
Sigala,~P.~A.; Fafarman,~A.~T.; Schwans,~J.~P.; Fried,~S.~D.; Fenn,~T.~D.;
  Caaveiro,~J. M.~M.; Pybus,~B.; Ringe,~D.; Petsko,~G.~A.; Boxer,~S.~G.
  \latin{et~al.}  Quantitative dissection of hydrogen bond-mediated proton
  transfer in the ketosteroid isomerase active site. \emph{Proc. Natl. Acad.
  Sci. U.S.A.} \textbf{2013}, \emph{110}, E2552--E2561\relax
\mciteBstWouldAddEndPuncttrue
\mciteSetBstMidEndSepPunct{\mcitedefaultmidpunct}
{\mcitedefaultendpunct}{\mcitedefaultseppunct}\relax
\EndOfBibitem
\bibitem[Pollack \latin{et~al.}(1989)Pollack, Zeng, Mack, and
  Eldin]{Pollackl1989}
Pollack,~R.~M.; Zeng,~B.; Mack,~J. P.~G.; Eldin,~S. Determination of the
  microscopic rate constants for the base catalyzed conjugation of
  5-androstene-3,17-dione. \emph{J. Am. Chem. Soc.} \textbf{1989}, \emph{111},
  6419--6423\relax
\mciteBstWouldAddEndPuncttrue
\mciteSetBstMidEndSepPunct{\mcitedefaultmidpunct}
{\mcitedefaultendpunct}{\mcitedefaultseppunct}\relax
\EndOfBibitem
\bibitem[Hawkinson \latin{et~al.}(1991)Hawkinson, Eames, and
  Pollack]{Hawkinson1991}
Hawkinson,~D.~C.; Eames,~T. C.~M.; Pollack,~R.~M. Energetics of
  3-oxo-$\Delta^5$-steroid isomerase: source of the catalytic power of the
  enzyme. \emph{Biochemistry} \textbf{1991}, \emph{30}, 10849--10858\relax
\mciteBstWouldAddEndPuncttrue
\mciteSetBstMidEndSepPunct{\mcitedefaultmidpunct}
{\mcitedefaultendpunct}{\mcitedefaultseppunct}\relax
\EndOfBibitem
\bibitem[Radzicka and Wolfenden(1995)Radzicka, and Wolfenden]{Radzicka1995}
Radzicka,~A.; Wolfenden,~R. A proficient enzyme. \emph{Science} \textbf{1995},
  \emph{267}, 90--93\relax
\mciteBstWouldAddEndPuncttrue
\mciteSetBstMidEndSepPunct{\mcitedefaultmidpunct}
{\mcitedefaultendpunct}{\mcitedefaultseppunct}\relax
\EndOfBibitem
\bibitem[Pollack(2004)]{Pollack2004}
Pollack,~R.~M. Enzymatic mechanisms for catalysis of enolization: ketosteroid
  isomerase. \emph{Bioorg. Chem.} \textbf{2004}, \emph{32}, 341--353\relax
\mciteBstWouldAddEndPuncttrue
\mciteSetBstMidEndSepPunct{\mcitedefaultmidpunct}
{\mcitedefaultendpunct}{\mcitedefaultseppunct}\relax
\EndOfBibitem
\bibitem[Warshel \latin{et~al.}(2007)Warshel, Sharma, Chu, and
  {\AA}qvist]{Warshel2007}
Warshel,~A.; Sharma,~P.~K.; Chu,~Z.~T.; {\AA}qvist,~J. Electrostatic
  Contributions to Binding of Transition State Analogues Can Be Very Different
  from the Corresponding Contributions to Catalysis: Phenolates Binding to the
  Oxyanion Hole of Ketosteroid Isomerase. \emph{Biochemistry} \textbf{2007},
  \emph{46}, 1466--1476\relax
\mciteBstWouldAddEndPuncttrue
\mciteSetBstMidEndSepPunct{\mcitedefaultmidpunct}
{\mcitedefaultendpunct}{\mcitedefaultseppunct}\relax
\EndOfBibitem
\bibitem[Kamerlin \latin{et~al.}(2010)Kamerlin, Sharma, Chu, and
  Warshel]{Kamerlin2010}
Kamerlin,~S.~C.; Sharma,~P.~K.; Chu,~Z.~T.; Warshel,~A. Ketosteroid isomerase
  provides further support for the idea that enzymes work by electrostatic
  preorganization. \emph{Proc. Natl. Acad. Sci. U.S.A.} \textbf{2010},
  \emph{107}, 4075--4080\relax
\mciteBstWouldAddEndPuncttrue
\mciteSetBstMidEndSepPunct{\mcitedefaultmidpunct}
{\mcitedefaultendpunct}{\mcitedefaultseppunct}\relax
\EndOfBibitem
\bibitem[Chakravorty and Hammes-Schiffer(2010)Chakravorty, and
  Hammes-Schiffer]{Chakravorty2010}
Chakravorty,~D.~K.; Hammes-Schiffer,~S. Impact of Mutation on Proton Transfer
  Reactions in Ketosteroid Isomerase: Insights from Molecular Dynamics
  Simulations. \emph{J. Am. Chem. Soc.} \textbf{2010}, \emph{132},
  7549--7555\relax
\mciteBstWouldAddEndPuncttrue
\mciteSetBstMidEndSepPunct{\mcitedefaultmidpunct}
{\mcitedefaultendpunct}{\mcitedefaultseppunct}\relax
\EndOfBibitem
\bibitem[Herschlag and Natarajan(2013)Herschlag, and Natarajan]{Herschlag2013}
Herschlag,~D.; Natarajan,~A. Fundamental Challenges in Mechanistic Enzymology:
  Progress toward Understanding the Rate Enhancements of Enzymes.
  \emph{Biochemistry} \textbf{2013}, \emph{52}, 2050--2067\relax
\mciteBstWouldAddEndPuncttrue
\mciteSetBstMidEndSepPunct{\mcitedefaultmidpunct}
{\mcitedefaultendpunct}{\mcitedefaultseppunct}\relax
\EndOfBibitem
\bibitem[Zhao \latin{et~al.}(1996)Zhao, Abeygunawardana, Talalay, and
  Mildvan]{Zhao1996}
Zhao,~Q.; Abeygunawardana,~C.; Talalay,~P.; Mildvan,~A.~S. NMR evidence for the
  participation of a low-barrier hydrogen bond in the mechanism of
  $\Delta^5$-3-ketosteroid isomerase. \emph{Proc. Natl. Acad. Sci. U.S.A.}
  \textbf{1996}, \emph{93}, 8220--8224\relax
\mciteBstWouldAddEndPuncttrue
\mciteSetBstMidEndSepPunct{\mcitedefaultmidpunct}
{\mcitedefaultendpunct}{\mcitedefaultseppunct}\relax
\EndOfBibitem
\bibitem[Childs and Boxer(2010)Childs, and Boxer]{Childs2010}
Childs,~W.; Boxer,~S.~G. Proton Affinity of the Oxyanion Hole in the Active
  Site of Ketosteroid Isomerase. \emph{Biochemistry} \textbf{2010}, \emph{49},
  2725--2731\relax
\mciteBstWouldAddEndPuncttrue
\mciteSetBstMidEndSepPunct{\mcitedefaultmidpunct}
{\mcitedefaultendpunct}{\mcitedefaultseppunct}\relax
\EndOfBibitem
\bibitem[Wu \latin{et~al.}(2015)Wu, Fried, and Boxer]{Wu2015}
Wu,~Y.; Fried,~S.~D.; Boxer,~S.~G. Dissecting proton delocalization in an
  enzyme's hydrogen bond network with unnatural amino acids.
  \emph{Biochemistry} \textbf{2015}, \emph{54}, 7110--7119\relax
\mciteBstWouldAddEndPuncttrue
\mciteSetBstMidEndSepPunct{\mcitedefaultmidpunct}
{\mcitedefaultendpunct}{\mcitedefaultseppunct}\relax
\EndOfBibitem
\bibitem[Wolfsberg \latin{et~al.}(2009)Wolfsberg, Van~Hook, Paneth, and
  Rebelo]{Wolfsberg2009}
Wolfsberg,~M.; Van~Hook,~W.; Paneth,~P.; Rebelo,~L. P.~N. \emph{Isotope
  Effects: in the Chemical, Geological, and Bio Sciences}; Springer: Dordrecht,
  2009\relax
\mciteBstWouldAddEndPuncttrue
\mciteSetBstMidEndSepPunct{\mcitedefaultmidpunct}
{\mcitedefaultendpunct}{\mcitedefaultseppunct}\relax
\EndOfBibitem
\bibitem[Hanoian \latin{et~al.}(2010)Hanoian, Sigala, Herschlag, and
  Hammes-Schiffer]{Hanoian2010}
Hanoian,~P.; Sigala,~P.~A.; Herschlag,~D.; Hammes-Schiffer,~S. Hydrogen Bonding
  in the Active Site of Ketosteroid Isomerase: Electronic Inductive Effects and
  Hydrogen Bond Coupling. \emph{Biochemistry} \textbf{2010}, \emph{49},
  10339--10348\relax
\mciteBstWouldAddEndPuncttrue
\mciteSetBstMidEndSepPunct{\mcitedefaultmidpunct}
{\mcitedefaultendpunct}{\mcitedefaultseppunct}\relax
\EndOfBibitem
\bibitem[Wang \latin{et~al.}(2014)Wang, Fried, Boxer, and Markland]{Wang2014a}
Wang,~L.; Fried,~S.~D.; Boxer,~S.~G.; Markland,~T.~E. Quantum delocalization of
  protons in the hydrogen-bond network of an enzyme active site. \emph{Proc.
  Natl. Acad. Sci. U.S.A.} \textbf{2014}, \emph{111}, 18454--18459\relax
\mciteBstWouldAddEndPuncttrue
\mciteSetBstMidEndSepPunct{\mcitedefaultmidpunct}
{\mcitedefaultendpunct}{\mcitedefaultseppunct}\relax
\EndOfBibitem
\bibitem[Schwans \latin{et~al.}(2013)Schwans, Sunden, Gonzalez, Tsai, and
  Herschlag]{Schwans2013}
Schwans,~J.~P.; Sunden,~F.; Gonzalez,~A.; Tsai,~Y.; Herschlag,~D. Uncovering
  the Determinants of a Highly Perturbed Tyrosine pKa in the Active Site of
  Ketosteroid Isomerase. \emph{Biochemistry} \textbf{2013}, \emph{52},
  7840--7855\relax
\mciteBstWouldAddEndPuncttrue
\mciteSetBstMidEndSepPunct{\mcitedefaultmidpunct}
{\mcitedefaultendpunct}{\mcitedefaultseppunct}\relax
\EndOfBibitem
\bibitem[Natarajan \latin{et~al.}(2014)Natarajan, Schwans, and
  Herschlag]{Natarajan2014}
Natarajan,~A.; Schwans,~J.~P.; Herschlag,~D. Using Unnatural Amino Acids to
  Probe the Energetics of Oxyanion Hole Hydrogen Bonds in the Ketosteroid
  Isomerase Active Site. \emph{J. Am. Chem. Soc.} \textbf{2014}, \emph{136},
  7643--7654\relax
\mciteBstWouldAddEndPuncttrue
\mciteSetBstMidEndSepPunct{\mcitedefaultmidpunct}
{\mcitedefaultendpunct}{\mcitedefaultseppunct}\relax
\EndOfBibitem
\bibitem[Fried \latin{et~al.}(2014)Fried, Bagchi, and Boxer]{Fried2014}
Fried,~S.~D.; Bagchi,~S.; Boxer,~S.~G. Extreme electric fields power catalysis
  in the active site of ketosteroid isomerase. \emph{Science} \textbf{2014},
  \emph{346}, 1510--1514\relax
\mciteBstWouldAddEndPuncttrue
\mciteSetBstMidEndSepPunct{\mcitedefaultmidpunct}
{\mcitedefaultendpunct}{\mcitedefaultseppunct}\relax
\EndOfBibitem
\bibitem[Wu and Boxer(2016)Wu, and Boxer]{Wu2016}
Wu,~Y.; Boxer,~S.~G. A Critical Test of the Electrostatic Contribution to
  Catalysis with Noncanonical Amino Acids in Ketosteroid Isomerase. \emph{J.
  Am. Chem. Soc.} \textbf{2016}, \emph{138}, 11890--11895\relax
\mciteBstWouldAddEndPuncttrue
\mciteSetBstMidEndSepPunct{\mcitedefaultmidpunct}
{\mcitedefaultendpunct}{\mcitedefaultseppunct}\relax
\EndOfBibitem
\bibitem[Lamba \latin{et~al.}(2016)Lamba, Yabukarski, Pinney, and
  Herschlag]{Lamba2016}
Lamba,~V.; Yabukarski,~F.; Pinney,~M.; Herschlag,~D. Evaluation of the
  Catalytic Contribution from a Positioned General Base in Ketosteroid
  Isomerase. \emph{J. Am. Chem. Soc.} \textbf{2016}, \emph{138},
  9902--9909\relax
\mciteBstWouldAddEndPuncttrue
\mciteSetBstMidEndSepPunct{\mcitedefaultmidpunct}
{\mcitedefaultendpunct}{\mcitedefaultseppunct}\relax
\EndOfBibitem
\bibitem[Fried and Boxer(2017)Fried, and Boxer]{Fried2017}
Fried,~S.~D.; Boxer,~S.~G. Electric Fields and Enzyme Catalysis. \emph{Annu.
  Rev. Biochem.} \textbf{2017}, \emph{86}, 387--415\relax
\mciteBstWouldAddEndPuncttrue
\mciteSetBstMidEndSepPunct{\mcitedefaultmidpunct}
{\mcitedefaultendpunct}{\mcitedefaultseppunct}\relax
\EndOfBibitem
\bibitem[Bhowmick \latin{et~al.}(2017)Bhowmick, Sharma, and
  Head-Gordon]{Bhowmick2017}
Bhowmick,~A.; Sharma,~S.~C.; Head-Gordon,~T. The Importance of the Scaffold for
  de Novo Enzymes: A Case Study with Kemp Eliminase. \emph{J. Am. Chem. Soc.}
  \textbf{2017}, \emph{139}, 5793--5800\relax
\mciteBstWouldAddEndPuncttrue
\mciteSetBstMidEndSepPunct{\mcitedefaultmidpunct}
{\mcitedefaultendpunct}{\mcitedefaultseppunct}\relax
\EndOfBibitem
\bibitem[Isborn \latin{et~al.}(2012)Isborn, G{\"{o}}tz, Clark, Walker, and
  Martinez]{Isborn2012}
Isborn,~C.~M.; G{\"{o}}tz,~A.~W.; Clark,~M.~A.; Walker,~R.~C.; Martinez,~T.~J.
  Electronic Absorption Spectra from MM and ab Initio QM/MM Molecular Dynamics:
  Environmental Effects on the Absorption Spectrum of Photoactive Yellow
  Protein. \emph{J. Chem. Theory Comput.} \textbf{2012}, \emph{8},
  5092--5106\relax
\mciteBstWouldAddEndPuncttrue
\mciteSetBstMidEndSepPunct{\mcitedefaultmidpunct}
{\mcitedefaultendpunct}{\mcitedefaultseppunct}\relax
\EndOfBibitem
\bibitem[Ufimtsev and Martinez(2009)Ufimtsev, and Martinez]{Ufimtsev2009}
Ufimtsev,~I.~S.; Martinez,~T.~J. Quantum Chemistry on Graphical Processing
  Units. 3. Analytical Energy Gradients, Geometry Optimization, and First
  Principles Molecular Dynamics. \emph{J. Chem. Theory Comput.} \textbf{2009},
  \emph{5}, 2619--2628\relax
\mciteBstWouldAddEndPuncttrue
\mciteSetBstMidEndSepPunct{\mcitedefaultmidpunct}
{\mcitedefaultendpunct}{\mcitedefaultseppunct}\relax
\EndOfBibitem
\bibitem[Chen \latin{et~al.}(2003)Chen, Ivanov, Klein, and
  Parrinello]{chen+03prl}
Chen,~B.; Ivanov,~I.; Klein,~M.~L.; Parrinello,~M. {Hydrogen Bonding in Water}.
  \emph{Phys. Rev. Lett.} \textbf{2003}, \emph{91}, 215503\relax
\mciteBstWouldAddEndPuncttrue
\mciteSetBstMidEndSepPunct{\mcitedefaultmidpunct}
{\mcitedefaultendpunct}{\mcitedefaultseppunct}\relax
\EndOfBibitem
\bibitem[Morrone and Car(2008)Morrone, and Car]{morr-car08prl}
Morrone,~J.~A.; Car,~R. {Nuclear Quantum Effects in Water}. \emph{Phys. Rev.
  Lett.} \textbf{2008}, \emph{101}, 17801\relax
\mciteBstWouldAddEndPuncttrue
\mciteSetBstMidEndSepPunct{\mcitedefaultmidpunct}
{\mcitedefaultendpunct}{\mcitedefaultseppunct}\relax
\EndOfBibitem
\bibitem[P{\'{e}}rez \latin{et~al.}(2010)P{\'{e}}rez, Tuckerman, Hjalmarson,
  and von Lilienfeld]{Perez2010}
P{\'{e}}rez,~A.; Tuckerman,~M.~E.; Hjalmarson,~H.~P.; von Lilienfeld,~O.~A.
  Enol Tautomers of Watson-Crick Base Pair Models Are Metastable Because of
  Nuclear Quantum Effects. \emph{J. Am. Chem. Soc.} \textbf{2010}, \emph{132},
  11510--11515\relax
\mciteBstWouldAddEndPuncttrue
\mciteSetBstMidEndSepPunct{\mcitedefaultmidpunct}
{\mcitedefaultendpunct}{\mcitedefaultseppunct}\relax
\EndOfBibitem
\bibitem[Li \latin{et~al.}(2011)Li, Walker, and Michaelides]{li+11pnas}
Li,~X.-Z.; Walker,~B.; Michaelides,~A. {Quantum nature of the hydrogen bond}.
  \emph{Proc. Natl. Acad. Sci. USA} \textbf{2011}, \emph{108}, 6369--6373\relax
\mciteBstWouldAddEndPuncttrue
\mciteSetBstMidEndSepPunct{\mcitedefaultmidpunct}
{\mcitedefaultendpunct}{\mcitedefaultseppunct}\relax
\EndOfBibitem
\bibitem[Markland and Berne(2012)Markland, and Berne]{mark-bern12pnas}
Markland,~T.~E.; Berne,~B.~J. {Unraveling quantum mechanical effects in water
  using isotopic fractionation}. \emph{Proc. Natl. Acad. Sci. U.S.A.}
  \textbf{2012}, \emph{109}, 7988--7991\relax
\mciteBstWouldAddEndPuncttrue
\mciteSetBstMidEndSepPunct{\mcitedefaultmidpunct}
{\mcitedefaultendpunct}{\mcitedefaultseppunct}\relax
\EndOfBibitem
\bibitem[Ceriotti \latin{et~al.}(2013)Ceriotti, Cuny, Parrinello, and
  Manolopoulos]{ceri+13pnas}
Ceriotti,~M.; Cuny,~J.; Parrinello,~M.; Manolopoulos,~D.~E. {Nuclear quantum
  effects and hydrogen bond fluctuations in water.} \emph{Proc. Natl. Acad.
  Sci. USA} \textbf{2013}, \emph{110}, 15591--6\relax
\mciteBstWouldAddEndPuncttrue
\mciteSetBstMidEndSepPunct{\mcitedefaultmidpunct}
{\mcitedefaultendpunct}{\mcitedefaultseppunct}\relax
\EndOfBibitem
\bibitem[Wang \latin{et~al.}(2014)Wang, Ceriotti, and Markland]{Wang2014}
Wang,~L.; Ceriotti,~M.; Markland,~T.~E. Quantum fluctuations and isotope
  effects in ab initio descriptions of water. \emph{J. Chem. Phys.}
  \textbf{2014}, \emph{141}, 104502\relax
\mciteBstWouldAddEndPuncttrue
\mciteSetBstMidEndSepPunct{\mcitedefaultmidpunct}
{\mcitedefaultendpunct}{\mcitedefaultseppunct}\relax
\EndOfBibitem
\bibitem[McKenzie \latin{et~al.}(2015)McKenzie, Athokpam, and
  Ramesh]{McKenzie2015}
McKenzie,~R.~H.; Athokpam,~B.; Ramesh,~S.~G. Isotopic fractionation in proteins
  as a measure of hydrogen bond length. \emph{J. Chem. Phys.} \textbf{2015},
  \emph{143}, 044309--\relax
\mciteBstWouldAddEndPuncttrue
\mciteSetBstMidEndSepPunct{\mcitedefaultmidpunct}
{\mcitedefaultendpunct}{\mcitedefaultseppunct}\relax
\EndOfBibitem
\bibitem[Rossi \latin{et~al.}(2015)Rossi, Fang, and Michaelides]{Rossi2015}
Rossi,~M.; Fang,~W.; Michaelides,~A. Stability of Complex Biomolecular
  Structures: van der Waals, Hydrogen Bond Cooperativity, and Nuclear Quantum
  Effects. \emph{J. Phys. Chem. Lett.} \textbf{2015}, \emph{6},
  4233--4238\relax
\mciteBstWouldAddEndPuncttrue
\mciteSetBstMidEndSepPunct{\mcitedefaultmidpunct}
{\mcitedefaultendpunct}{\mcitedefaultseppunct}\relax
\EndOfBibitem
\bibitem[Ceriotti \latin{et~al.}(2016)Ceriotti, Fang, Kusalik, McKenzie,
  Michaelides, Morales, and Markland]{Ceriotti2016}
Ceriotti,~M.; Fang,~W.; Kusalik,~P.~G.; McKenzie,~R.~H.; Michaelides,~A.;
  Morales,~M.~A.; Markland,~T.~E. Nuclear Quantum Effects in Water and Aqueous
  Systems: Experiment, Theory, and Current Challenges. \emph{Chem. Rev.}
  \textbf{2016}, \emph{116}, 7529--7550\relax
\mciteBstWouldAddEndPuncttrue
\mciteSetBstMidEndSepPunct{\mcitedefaultmidpunct}
{\mcitedefaultendpunct}{\mcitedefaultseppunct}\relax
\EndOfBibitem
\bibitem[Feynman and Hibbs(1964)Feynman, and Hibbs]{feyn-hibb65book}
Feynman,~R.~P.; Hibbs,~A.~R. \emph{{Quantum Mechanics and Path Integrals}};
  McGraw-Hill: New York, 1964\relax
\mciteBstWouldAddEndPuncttrue
\mciteSetBstMidEndSepPunct{\mcitedefaultmidpunct}
{\mcitedefaultendpunct}{\mcitedefaultseppunct}\relax
\EndOfBibitem
\bibitem[Chandler and Wolynes(1981)Chandler, and Wolynes]{chan-woly81jcp}
Chandler,~D.; Wolynes,~P.~G. {Exploiting the isomorphism between quantum theory
  and classical statistical mechanics of polyatomic fluids}. \emph{J. Chem.
  Phys.} \textbf{1981}, \emph{74}, 4078--4095\relax
\mciteBstWouldAddEndPuncttrue
\mciteSetBstMidEndSepPunct{\mcitedefaultmidpunct}
{\mcitedefaultendpunct}{\mcitedefaultseppunct}\relax
\EndOfBibitem
\bibitem[Berne and Thirumalai(1986)Berne, and Thirumalai]{bern-thir86arpc}
Berne,~B.~J.; Thirumalai,~D. {On the Simulation of Quantum Systems: Path
  Integral Methods}. \emph{Ann. Rev. Phys. Chem.} \textbf{1986}, \emph{37},
  401--424\relax
\mciteBstWouldAddEndPuncttrue
\mciteSetBstMidEndSepPunct{\mcitedefaultmidpunct}
{\mcitedefaultendpunct}{\mcitedefaultseppunct}\relax
\EndOfBibitem
\bibitem[Wang \latin{et~al.}(2016)Wang, Isborn, and Markland]{Wang2016}
Wang,~L.; Isborn,~C.; Markland,~T. Simulating Nuclear and Electronic Quantum
  Effects in Enzymes. \emph{Methods Enzymol.} \textbf{2016}, \emph{577},
  389 -- 418\relax
\mciteBstWouldAddEndPuncttrue
\mciteSetBstMidEndSepPunct{\mcitedefaultmidpunct}
{\mcitedefaultendpunct}{\mcitedefaultseppunct}\relax
\EndOfBibitem
\bibitem[Frisch \latin{et~al.}()Frisch, Trucks, Schlegel, Scuseria, Robb,
  Cheeseman, Scalmani, Barone, Mennucci, Petersson, Nakatsuji, Caricato, Li,
  Hratchian, Izmaylov, Bloino, Zheng, Sonnenberg, Hada, Ehara, Toyota, Fukuda,
  Hasegawa, Ishida, Nakajima, Honda, Kitao, Nakai, Vreven, Montgomery, Peralta,
  Ogliaro, Bearpark, Heyd, Brothers, Kudin, Staroverov, Kobayashi, Normand,
  Raghavachari, Rendell, Burant, Iyengar, Tomasi, Cossi, Rega, Millam, Klene,
  Knox, Cross, Bakken, Adamo, Jaramillo, Gomperts, Stratmann, Yazyev, Austin,
  Cammi, Pomelli, Ochterski, Martin, Morokuma, Zakrzewski, Voth, Salvador,
  Dannenberg, Dapprich, Daniels, Farkas, Foresman, Ortiz, Cioslowski, and
  Fox]{g09}
Frisch,~M.~J.; Trucks,~G.~W.; Schlegel,~H.~B.; Scuseria,~G.~E.; Robb,~M.~A.;
  Cheeseman,~J.~R.; Scalmani,~G.; Barone,~V.; Mennucci,~B.; Petersson,~G.~A.
  \latin{et~al.}  Gaussian09 {R}evision {D}.01\relax
\mciteBstWouldAddEndPuncttrue
\mciteSetBstMidEndSepPunct{\mcitedefaultmidpunct}
{\mcitedefaultendpunct}{\mcitedefaultseppunct}\relax
\EndOfBibitem
\bibitem[Becke(1993)]{beck93jcp}
Becke,~A.~D. {Density-functional thermochemistry. III. The role of exact
  exchange}. \emph{J. Chem. Phys.} \textbf{1993}, \emph{98}, 5648\relax
\mciteBstWouldAddEndPuncttrue
\mciteSetBstMidEndSepPunct{\mcitedefaultmidpunct}
{\mcitedefaultendpunct}{\mcitedefaultseppunct}\relax
\EndOfBibitem
\bibitem[Duan \latin{et~al.}(2003)Duan, Wu, Chowdhury, Lee, Xiong, Zhang, Yang,
  Cieplak, Luo, Lee, Caldwell, Wang, and Kollman]{Duan2003}
Duan,~Y.; Wu,~C.; Chowdhury,~S.; Lee,~M.~C.; Xiong,~G.; Zhang,~W.; Yang,~R.;
  Cieplak,~P.; Luo,~R.; Lee,~T. \latin{et~al.}  A point-charge force field for
  molecular mechanics simulations of proteins based on condensed-phase quantum
  mechanical calculations. \emph{J. Comput. Chem.} \textbf{2003}, \emph{24},
  1999--2012\relax
\mciteBstWouldAddEndPuncttrue
\mciteSetBstMidEndSepPunct{\mcitedefaultmidpunct}
{\mcitedefaultendpunct}{\mcitedefaultseppunct}\relax
\EndOfBibitem
\bibitem[Jorgensen \latin{et~al.}(1983)Jorgensen, Chandrasekhar, Madura, Impey,
  and Klein]{jorg+83jcp}
Jorgensen,~W. W.~L.; Chandrasekhar,~J.; Madura,~J.~D.; Impey,~R.~W.;
  Klein,~M.~L. {Comparison of simple potential functions for simulating liquid
  water}. \emph{J. Chem. Phys.} \textbf{1983}, \emph{79}, 926\relax
\mciteBstWouldAddEndPuncttrue
\mciteSetBstMidEndSepPunct{\mcitedefaultmidpunct}
{\mcitedefaultendpunct}{\mcitedefaultseppunct}\relax
\EndOfBibitem
\bibitem[Habershon \latin{et~al.}(2009)Habershon, Markland, and
  Manolopoulos]{habe+09jcp}
Habershon,~S.; Markland,~T.~E.; Manolopoulos,~D.~E. {Competing quantum effects
  in the dynamics of a flexible water model.} \emph{J. Chem. Phys.}
  \textbf{2009}, \emph{131}, 24501\relax
\mciteBstWouldAddEndPuncttrue
\mciteSetBstMidEndSepPunct{\mcitedefaultmidpunct}
{\mcitedefaultendpunct}{\mcitedefaultseppunct}\relax
\EndOfBibitem
\bibitem[Romanelli \latin{et~al.}(2013)Romanelli, Ceriotti, Manolopoulos,
  Pantalei, Senesi, and Andreani]{roma+13jpcl}
Romanelli,~G.; Ceriotti,~M.; Manolopoulos,~D.~E.; Pantalei,~C.; Senesi,~R.;
  Andreani,~C. {Direct Measurement of Competing Quantum Effects on the Kinetic
  Energy of Heavy Water upon Melting}. \emph{J. Phys. Chem. Letters}
  \textbf{2013}, \emph{4}, 3251--3256\relax
\mciteBstWouldAddEndPuncttrue
\mciteSetBstMidEndSepPunct{\mcitedefaultmidpunct}
{\mcitedefaultendpunct}{\mcitedefaultseppunct}\relax
\EndOfBibitem
\bibitem[Fang \latin{et~al.}(2016)Fang, Chen, Rossi, Feng, Li, and
  Michaelides]{Fang2016}
Fang,~W.; Chen,~J.; Rossi,~M.; Feng,~Y.; Li,~X.-Z.; Michaelides,~A. Inverse
  Temperature Dependence of Nuclear Quantum Effects in DNA Base Pairs. \emph{J.
  Phys. Chem. Lett.} \textbf{2016}, \emph{7}, 2125--2131\relax
\mciteBstWouldAddEndPuncttrue
\mciteSetBstMidEndSepPunct{\mcitedefaultmidpunct}
{\mcitedefaultendpunct}{\mcitedefaultseppunct}\relax
\EndOfBibitem
\bibitem[Shi \latin{et~al.}(2015)Shi, Shin, Hassanali, and Singer]{Shi2015}
Shi,~B.; Shin,~Y.~K.; Hassanali,~A.~A.; Singer,~S.~J. DNA Binding to the Silica
  Surface. \emph{J. Phys. Chem. B} \textbf{2015}, \emph{119},
  11030--11040\relax
\mciteBstWouldAddEndPuncttrue
\mciteSetBstMidEndSepPunct{\mcitedefaultmidpunct}
{\mcitedefaultendpunct}{\mcitedefaultseppunct}\relax
\EndOfBibitem
\bibitem[Kim \latin{et~al.}(2000)Kim, Jang, Nam, Choi, Kim, Ha, Kim, Oh, and
  Choi]{Kim2000}
Kim,~D.-H.; Jang,~D.~S.; Nam,~G.~H.; Choi,~G.; Kim,~J.-S.; Ha,~N.-C.;
  Kim,~M.-S.; Oh,~B.-H.; Choi,~K.~Y. Contribution of the Hydrogen-Bond Network
  Involving a Tyrosine Triad in the Active Site to the Structure and Function
  of a Highly Proficient Ketosteroid Isomerase from Pseudomonas putida Biotype
  B,. \emph{Biochemistry} \textbf{2000}, \emph{39}, 4581--4589\relax
\mciteBstWouldAddEndPuncttrue
\mciteSetBstMidEndSepPunct{\mcitedefaultmidpunct}
{\mcitedefaultendpunct}{\mcitedefaultseppunct}\relax
\EndOfBibitem
\bibitem[Choi \latin{et~al.}(2001)Choi, Ha, Kim, Hong, Oh, and Choi]{Choi2001}
Choi,~G.; Ha,~N.-C.; Kim,~M.-S.; Hong,~B.-H.; Oh,~B.-H.; Choi,~K.~Y.
  Pseudoreversion of the Catalytic Activity of Y14F by the Additional
  Substitution(s) of Tyrosine with Phenylalanine in the Hydrogen Bond Network
  of $\Delta^5$-3-Ketosteroid Isomerase from Pseudomonas putida Biotype B.
  \emph{Biochemistry} \textbf{2001}, \emph{40}, 6828--6835\relax
\mciteBstWouldAddEndPuncttrue
\mciteSetBstMidEndSepPunct{\mcitedefaultmidpunct}
{\mcitedefaultendpunct}{\mcitedefaultseppunct}\relax
\EndOfBibitem

\bibitem[]{Footnote}
We note that these mutation experiments do not simply isolate the contribution from the removed residue but rather replace it with another amino acid, and thus the measured electric fields might also have contributions from the residues they are mutated to.
\mciteBstWouldAddEndPuncttrue
\mciteSetBstMidEndSepPunct{\mcitedefaultmidpunct}
{\mcitedefaultendpunct}{\mcitedefaultseppunct}\relax
\EndOfBibitem

\end{mcitethebibliography}

\begin{mcitethebibliography}{16}
\providecommand*\natexlab[1]{#1}
\providecommand*\mciteSetBstSublistMode[1]{}
\providecommand*\mciteSetBstMaxWidthForm[2]{}
\providecommand*\mciteBstWouldAddEndPuncttrue
  {\def\EndOfBibitem{\unskip.}}
\providecommand*\mciteBstWouldAddEndPunctfalse
  {\let\EndOfBibitem\relax}
\providecommand*\mciteSetBstMidEndSepPunct[3]{}
\providecommand*\mciteSetBstSublistLabelBeginEnd[3]{}
\providecommand*\EndOfBibitem{}
\mciteSetBstSublistMode{f}
\mciteSetBstMaxWidthForm{subitem}{(\alph{mcitesubitemcount})}
\mciteSetBstSublistLabelBeginEnd
  {\mcitemaxwidthsubitemform\space}
  {\relax}
  {\relax}

\bibitem[Wang \latin{et~al.}(2014)Wang, Fried, Boxer, and Markland]{Wang2014a}
Wang,~L.; Fried,~S.~D.; Boxer,~S.~G.; Markland,~T.~E. Quantum delocalization of
  protons in the hydrogen-bond network of an enzyme active site. \emph{Proc.
  Natl. Acad. Sci. USA} \textbf{2014}, \emph{111}, 18454--18459\relax
\mciteBstWouldAddEndPuncttrue
\mciteSetBstMidEndSepPunct{\mcitedefaultmidpunct}
{\mcitedefaultendpunct}{\mcitedefaultseppunct}\relax
\EndOfBibitem
\bibitem[Becke(1993)]{beck93jcp}
Becke,~A.~D. {Density-functional thermochemistry. III. The role of exact
  exchange}. \emph{J. Chem. Phys.} \textbf{1993}, \emph{98}, 5648\relax
\mciteBstWouldAddEndPuncttrue
\mciteSetBstMidEndSepPunct{\mcitedefaultmidpunct}
{\mcitedefaultendpunct}{\mcitedefaultseppunct}\relax
\EndOfBibitem
\bibitem[Grimme \latin{et~al.}(2010)Grimme, Antony, Ehrlich, and
  Krieg]{grim+10jcp}
Grimme,~S.; Antony,~J.; Ehrlich,~S.; Krieg,~H. {A consistent and accurate ab
  initio parametrization of density functional dispersion correction (DFT-D)
  for the 94 elements H-Pu.} \emph{J. Chem. Phys.} \textbf{2010}, \emph{132},
  154104\relax
\mciteBstWouldAddEndPuncttrue
\mciteSetBstMidEndSepPunct{\mcitedefaultmidpunct}
{\mcitedefaultendpunct}{\mcitedefaultseppunct}\relax
\EndOfBibitem
\bibitem[Ceriotti and Manolopoulos(2012)Ceriotti, and
  Manolopoulos]{ceri-mano12prl}
Ceriotti,~M.; Manolopoulos,~D.~E. {Efficient First-Principles Calculation of
  the Quantum Kinetic Energy and Momentum Distribution of Nuclei}. \emph{Phys.
  Rev. Lett.} \textbf{2012}, \emph{109}, 100604\relax
\mciteBstWouldAddEndPuncttrue
\mciteSetBstMidEndSepPunct{\mcitedefaultmidpunct}
{\mcitedefaultendpunct}{\mcitedefaultseppunct}\relax
\EndOfBibitem
\bibitem[Duan \latin{et~al.}(2003)Duan, Wu, Chowdhury, Lee, Xiong, Zhang, Yang,
  Cieplak, Luo, Lee, Caldwell, Wang, and Kollman]{Duan2003}
Duan,~Y.; Wu,~C.; Chowdhury,~S.; Lee,~M.~C.; Xiong,~G.; Zhang,~W.; Yang,~R.;
  Cieplak,~P.; Luo,~R.; Lee,~T. \latin{et~al.}  A point-charge force field for
  molecular mechanics simulations of proteins based on condensed-phase quantum
  mechanical calculations. \emph{J. Comput. Chem.} \textbf{2003}, \emph{24},
  1999--2012\relax
\mciteBstWouldAddEndPuncttrue
\mciteSetBstMidEndSepPunct{\mcitedefaultmidpunct}
{\mcitedefaultendpunct}{\mcitedefaultseppunct}\relax
\EndOfBibitem
\bibitem[Jorgensen \latin{et~al.}(1983)Jorgensen, Chandrasekhar, Madura, Impey,
  and Klein]{jorg+83jcp}
Jorgensen,~W. W.~L.; Chandrasekhar,~J.; Madura,~J.~D.; Impey,~R.~W.;
  Klein,~M.~L. {Comparison of simple potential functions for simulating liquid
  water}. \emph{J. Chem. Phys.} \textbf{1983}, \emph{79}, 926\relax
\mciteBstWouldAddEndPuncttrue
\mciteSetBstMidEndSepPunct{\mcitedefaultmidpunct}
{\mcitedefaultendpunct}{\mcitedefaultseppunct}\relax
\EndOfBibitem
\bibitem[Ufimtsev and Martinez(2009)Ufimtsev, and Martinez]{Ufimtsev2009}
Ufimtsev,~I.~S.; Martinez,~T.~J. Quantum Chemistry on Graphical Processing
  Units. 3. Analytical Energy Gradients, Geometry Optimization, and First
  Principles Molecular Dynamics. \emph{J. Chem. Theory Comput.} \textbf{2009},
  \emph{5}, 2619--2628\relax
\mciteBstWouldAddEndPuncttrue
\mciteSetBstMidEndSepPunct{\mcitedefaultmidpunct}
{\mcitedefaultendpunct}{\mcitedefaultseppunct}\relax
\EndOfBibitem
\bibitem[Isborn \latin{et~al.}(2012)Isborn, G{\"{o}}tz, Clark, Walker, and
  Martinez]{Isborn2012}
Isborn,~C.~M.; G{\"{o}}tz,~A.~W.; Clark,~M.~A.; Walker,~R.~C.; Martinez,~T.~J.
  Electronic Absorption Spectra from MM and ab Initio QM/MM Molecular Dynamics:
  Environmental Effects on the Absorption Spectrum of Photoactive Yellow
  Protein. \emph{J. Chem. Theory Comput.} \textbf{2012}, \emph{8},
  5092--5106\relax
\mciteBstWouldAddEndPuncttrue
\mciteSetBstMidEndSepPunct{\mcitedefaultmidpunct}
{\mcitedefaultendpunct}{\mcitedefaultseppunct}\relax
\EndOfBibitem
\bibitem[Plimpton(1995)]{plim95jcp}
Plimpton,~S. {Fast Parallel Algorithms for Short-Range Molecular Dynamics}.
  \emph{J. Comp. Phys.} \textbf{1995}, \emph{117}, 1--19\relax
\mciteBstWouldAddEndPuncttrue
\mciteSetBstMidEndSepPunct{\mcitedefaultmidpunct}
{\mcitedefaultendpunct}{\mcitedefaultseppunct}\relax
\EndOfBibitem
\bibitem[Ha \latin{et~al.}(2000)Ha, Kim, Lee, Choi, and Oh]{Ha2000}
Ha,~N.-C.; Kim,~M.-S.; Lee,~W.; Choi,~K.~Y.; Oh,~B.-H. Detection of Large pKa
  Perturbations of an Inhibitor and a Catalytic Group at an Enzyme Active Site,
  a Mechanistic Basis for Catalytic Power of Many Enzymes. \emph{J. Biol.
  Chem.} \textbf{2000}, \emph{275}, 41100--41106\relax
\mciteBstWouldAddEndPuncttrue
\mciteSetBstMidEndSepPunct{\mcitedefaultmidpunct}
{\mcitedefaultendpunct}{\mcitedefaultseppunct}\relax
\EndOfBibitem
\bibitem[Frisch \latin{et~al.}()Frisch, Trucks, Schlegel, Scuseria, Robb,
  Cheeseman, Scalmani, Barone, Mennucci, Petersson, Nakatsuji, Caricato, Li,
  Hratchian, Izmaylov, Bloino, Zheng, Sonnenberg, Hada, Ehara, Toyota, Fukuda,
  Hasegawa, Ishida, Nakajima, Honda, Kitao, Nakai, Vreven, Montgomery, Peralta,
  Ogliaro, Bearpark, Heyd, Brothers, Kudin, Staroverov, Kobayashi, Normand,
  Raghavachari, Rendell, Burant, Iyengar, Tomasi, Cossi, Rega, Millam, Klene,
  Knox, Cross, Bakken, Adamo, Jaramillo, Gomperts, Stratmann, Yazyev, Austin,
  Cammi, Pomelli, Ochterski, Martin, Morokuma, Zakrzewski, Voth, Salvador,
  Dannenberg, Dapprich, Daniels, Farkas, Foresman, Ortiz, Cioslowski, and
  Fox]{g09}
Frisch,~M.~J.; Trucks,~G.~W.; Schlegel,~H.~B.; Scuseria,~G.~E.; Robb,~M.~A.;
  Cheeseman,~J.~R.; Scalmani,~G.; Barone,~V.; Mennucci,~B.; Petersson,~G.~A.
  \latin{et~al.}  Gaussian09 {R}evision {D}.01. \relax
\mciteBstWouldAddEndPuncttrue
\mciteSetBstMidEndSepPunct{\mcitedefaultmidpunct}
{\mcitedefaultendpunct}{\mcitedefaultseppunct}\relax
\EndOfBibitem
\bibitem[Ceriotti and Markland(2013)Ceriotti, and Markland]{ceri-mark13jcp}
Ceriotti,~M.; Markland,~T.~E. {Efficient methods and practical guidelines for
  simulating isotope effects.} \emph{J. Chem. Phys.} \textbf{2013}, \emph{138},
  014112\relax
\mciteBstWouldAddEndPuncttrue
\mciteSetBstMidEndSepPunct{\mcitedefaultmidpunct}
{\mcitedefaultendpunct}{\mcitedefaultseppunct}\relax
\EndOfBibitem
\bibitem[Herman \latin{et~al.}(1982)Herman, Bruskin, and Berne]{herm-bern82jcp}
Herman,~M.~F.; Bruskin,~E.~J.; Berne,~B.~J. {On path integral Monte Carlo
  simulations}. \emph{J. Chem. Phys.} \textbf{1982}, \emph{76}, 5150\relax
\mciteBstWouldAddEndPuncttrue
\mciteSetBstMidEndSepPunct{\mcitedefaultmidpunct}
{\mcitedefaultendpunct}{\mcitedefaultseppunct}\relax
\EndOfBibitem
\bibitem[Cao and Berne(1989)Cao, and Berne]{cao-bern89jcp}
Cao,~J.; Berne,~B.~J. {On energy estimators in path integral Monte Carlo
  simulations: Dependence of accuracy on algorithm}. \emph{J. Chem. Phys.}
  \textbf{1989}, \emph{91}, 6359\relax
\mciteBstWouldAddEndPuncttrue
\mciteSetBstMidEndSepPunct{\mcitedefaultmidpunct}
{\mcitedefaultendpunct}{\mcitedefaultseppunct}\relax
\EndOfBibitem
\end{mcitethebibliography}
\end{document}